\documentclass[usenatbib,usegraphicx]{mn2e}
\usepackage{float}
\usepackage{color}
\usepackage{amsmath}
\usepackage{enumerate}

\usepackage{times}

\usepackage{amsmath}
\usepackage{amssymb}
\usepackage{textcomp}

\newcommand{\bea}{\begin{eqnarray}}
\newcommand{\eea}{\end{eqnarray}}
\newcommand{\bc}{\begin{center}}
\newcommand{\ec}{\end{center}}

\renewcommand{\vec}[1]{ {\bmath #1} }

\newcommand{\dd}{{\rm d}}

\newcommand{\mach}{0.3}
\newcommand{\mtrans}{1.2}
\newcommand{\mmed}{3.5}
\newcommand{\mhigh}{8.4}

\renewcommand{\thefootnote}{\fnsymbol{footnote}}
\newcommand{\vect}[1]{\boldsymbol{#1}}

\title[Turbulence in SPH and the moving-mesh code AREPO]
{Subsonic turbulence in smoothed particle hydrodynamics and
  moving-mesh simulations} \author[A.~Bauer and V.~Springel]
{Andreas~Bauer$^1$\thanks{E-mail: andreas.bauer@h-its.org} and Volker~Springel$^{1,2}$\vspace{0.2cm}\\
$^1$Heidelberger Institut f\"{u}r Theoretische Studien,
  Schloss-Wolfsbrunnenweg 35, 69118 Heidelberg, Germany\\
$^2$Zentrum f\"ur Astronomie der Universit\"at Heidelberg,
  Astronomisches Recheninstitut, M\"{o}nchhofstr. 12-14, 69120
  Heidelberg, Germany}

\begin{document}
\maketitle

\begin{abstract}
  Highly supersonic, compressible turbulence is thought to be of
  tantamount importance for star formation processes in the
  interstellar medium (ISM). Likewise, cosmic structure formation is
  expected to give rise to subsonic turbulence in the intergalactic
  medium (IGM), which may substantially modify the thermodynamic
  structure of gas in virialized dark matter halos and affect
  small-scale mixing processes in the gas. Numerical simulations have
  played a key role in characterizing the properties of astrophysical
  turbulence, but thus far systematic code comparisons have been
  restricted to the supersonic regime, leaving it unclear whether
  subsonic turbulence is faithfully represented by the numerical
  techniques commonly employed in astrophysics. Here we focus on
  comparing the accuracy of smoothed particle hydrodynamics (SPH) and
  our new moving-mesh technique {\small AREPO} in simulations of
  driven subsonic turbulence. To make contact with previous results,
  we also analyze simulations of transsonic and highly supersonic
  turbulence. We find that the widely employed standard formulation of
  SPH yields problematic results in the subsonic regime. Instead of
  building up a Kolmogorov-like turbulent cascade, large-scale eddies
  are quickly damped close to the driving scale and decay into
  small-scale velocity noise. Reduced viscosity settings improve the
  situation, but the shape of the dissipation range differs
    compared with expectations for a Kolmogorov cascade.  In
  contrast, our moving-mesh technique does yield power-law scaling
  laws for the power spectra of velocity, vorticity and density,
  consistent with expectations for fully developed isotropic
  turbulence. We show that large errors in SPH's gradient estimate and
  the associated subsonic velocity noise are ultimately responsible
  for producing inaccurate results in the subsonic regime. In
  contrast, SPH's performance is much better for supersonic
  turbulence, as here the flow is kinetically dominated and
  characterized by a network of strong shocks, which can be adequately
  captured with SPH.  When compared to fixed-grid Eulerian simulations
  of turbulence, our moving-mesh approach shows qualitatively very
  similar results, although with somewhat better resolving power at
  the same number of cells, thanks to reduced advection errors and the
  automatic adaptivity of the {\small AREPO} code.
\end{abstract}

\section{Introduction}
\renewcommand{\thefootnote}{\fnsymbol{footnote}}

Astrophysical gas dynamics in the interstellar and intergalactic
medium is typically characterized by very high Reynolds numbers, thanks
to the comparatively low gas densities encountered in these
environments, which imply a very low physical viscosity for the
involved gas. We may hence expect that turbulent cascades over large
dynamic ranges are rather prevalent, provided effective driving
processes exist. Such turbulence can then be an important feature of
gas dynamics, for example providing an additional effective pressure
contribution, or leading to thorough small-scale mixing of chemical
elements in the gas.

In fact, it is believed that turbulence in the interstellar medium
(ISM) plays a key role in the formation of ordinary stellar
populations, determining in part the initial mass function of stars,
the lifetime of molecular clouds, and the overall efficiency of star
formation \citep[e.g.][]{Klessen2000, MacLow2004}. Here the turbulence
is highly supersonic, and presumably driven primarily by supernova
explosions. In addition, the strong radiative cooling processes of the
ISM make its equation-of-state approximately isothermal, such that
very strong shocks and high compression ratios are associated with the
supersonic gas motions. An additional complexity arises from magnetic
fields that are flux-frozen into the gas, such that the relevant
behaviour is that of isothermal, driven, supersonic,
magnetohydrodynamic turbulence.

Another regime where turbulence is thought to be important lies in
cosmological structure formation, particularly in virialized gaseous
halos, and in mildly non-linear filaments. Here small-scale random
motions may contribute a significant fraction of the pressure support
in group- and cluster-sized halos.  For example, \citet{Schuecker2004}
analysed pressure fluctuations in the Coma cluster, finding them to be
well described by a Kolmogorov power spectrum with a lower limit of
10\% of the total ICM pressure being in turbulent form.  Besides this
direct observational evidence, there are also strong analytic
arguments that suggest that hierarchical mergers should be able to
generate and sustain subsonic turbulence in galaxy clusters
\citep{Subramanian2006}.

Turbulence in the intracluster (ICM) and intergalactic medium (IGM) is
expected to be primarily of subsonic character, allowing it to be
approximately described as incompressible turbulence. The linear
growth of structure driven by gravity is not expected to be able to
generate this turbulence directly, due to the irrotational character
of the gravitational force field. However, during non-linear structure
formation, curved accretion and flow shocks can introduce vorticity
into the gas. Also, the baroclinic term in the wake of mergers can act
as an efficient source of vorticity, providing a large-scale driving
of intrahalo turbulence. Already a moderate degree of gas turbulence
may then have important consequences for the thermodynamic structure
of quasi hydrostatic halos. For example, turbulence may help to create
entropy cores in clusters \citep{Mitchell2009, ZuHone2011}, and its
dissipation provides for thermal heating. Also, the transport and the
mixing of metals is affected by turbulent gas motions. All of these
effects can modify the radiative cooling rates of intrahalo plasma,
which immediately impacts galaxy formation at the halo centers
\citep{Vogelsberger2011}. In addition, turbulence can be expected to
influence the magnetic field structure in virialized halo gas, which
in turn affects transport processes in the plasma such as thermal
conduction \citep[e.g.][]{Parrish2011} or physical viscosity
\citep{Sijacki2006}.

A considerable difficulty for the understanding of turbulence lies in
the comparatively limited quantitative knowledge that could thus far
be gained from purely analytic considerations. While Kolmogorov's
theory for the scaling laws of self-similar, incompressible turbulence
still stands out as one of the most insightful characterizations of
the physics of turbulence, an assessment of the accuracy of theories
for turbulence, especially when constructed for particularly
challenging cases such as compressible magnetohydrodynamic turbulence,
has relied to a large extent on numerical simulations. Most
simulations of astrophysical flows solve the Euler equations and not
the Navier-Stokes equations, based on the realization that the
residual physical viscosity is overwhelmed by numerical viscosity
anyway at the achievable resolutions. The dynamic range per dimension
that can be realized in 3D turbulence simulations is indeed quite
small, rarely exceeding a factor $10^3$ at present. Thus only a small
intertial range and comparatively low Reynolds numbers can be resolved
directly. Nevertheless, numerous numerical studies of the properties
of astrophysical turbulence have been carried out and have already led
to significant advances in our understanding of this important
phenomenon.

This is especially true for the study of highly supersonic turbulence
relevant for star formation, where a large body of literature has been
accumulated, including systematic comparisons of different numerical
techniques \citep{MacLow1998,Kitsionas2009,Kritsuk2011}.  For example,
\citet{Kitsionas2009} have  compared different
hydrodynamical codes (4 mesh codes and 3 SPH codes) when applied to
the decay of supersonic turbulence. They found generally somewhat
higher dissipation rates in SPH (which can however be modified with
different artificial viscosity parameterizations), but concluded that
on scales resolved with at least 32 resolution elements per dimension,
results were found to be qualitatively similar, at least as far as the
decay of the Mach number with time was concerned.  The velocity power
spectra, however, revealed that the damping on smaller scales was
consistently larger in SPH than in the grid codes.
 
In contrast, \citet{Price2010} claimed excellent agreement for driven
supersonic Mach ${\cal M}=10$ turbulence between the SPH-code {\small
  PHANTOM} and the Eulerian mesh-code {\small FLASH}.  In particular,
they found a consistent Kolmogorov-like slope in the power spectrum of
the variable $\rho^{1/3} v$. Their plain velocity power spectra also
agreed over an extended range of wave number $k$ at large scales,
although the SPH result eventually dipped down earlier. In the density
power spectrum on the other hand, their SPH result yielded more power
at high $k$, which can be interpreted as a welcome result of the
adaptive resolution of SPH. \citet{Price2010} also argued that
previous claims of a steeper slope for the turbulence spectrum in SPH
by \citet{Padoan2007}, based on simulations presented by
\citet{Ballesteros2006}, may have been a resolution effect only.

While the agreement reported by \citet{Price2010} for SPH and mesh
codes is encouraging, it is important to keep in mind that this was
achieved for highly supersonic turbulence. The very different physics
of such a flow compared with subsonic turbulence should caution
against taking it for granted that this reassuring success carries
over to subsonic flow as well.  In particular, we note that
\citet{Padoan2007} observed that there appears to be a Mach number
dependence of the turbulent slope in SPH, where for ${\cal M} =3$ a
slightly steeper (and hence `more wrong') slope was obtained
than for ${\cal M} =6$. This may mean that one tends to get a more
accurate result with SPH when the Mach number is high. Indeed, in a
response paper to our study, \citet{Price2012} pointed out that for
higher Mach number one naturally expects a higher Reynolds number in
SPH for given viscosity settings, which should then also allow a
larger inertial range.

In order to clarify these issues further, we focus in this paper on the
topic of subsonic turbulence, which is thought to be important in
cosmological structure formation.  For example, \citet{Ryu2008}
propose that vorticity generation along large-scale structure
formation shocks creates significant turbulence outside filaments, and
plays an important role in the production of intergalactic magnetic
fields in clusters, groups and filaments. \citet{Dolag2005} and
\citet{Vazza2006} used SPH to study turbulence in galaxy clusters
formed in cosmological simulations. They found evidence for scaling
laws where the total turbulent energy in the ICM scales with cluster
mass. \citet{Lau2009} argued that the effective pressure associated
with ICM turbulence may introduce a bias in the cluster masses
inferred from hydrostatic models.  \citet{Dolag2005} introduced and
tested a scheme for reduced viscosity in SPH, finding that this
produces significantly higher levels of turbulent gas motions than for
ordinary viscosity, reaching up to 5-30\% of the thermal energy. In
this study, they also measured a turbulent power spectrum for the
central $500\,{\rm kpc}$ of a simulated cluster, finding a
significantly shallower slope than expected for Kolmogorov turbulence.

Similar results were obtained by \citet{Valdarnini2011}, who has
presented the most detailed study of cluster turbulence based on SPH
thus far, including an investigation of the impact of artificial
viscosity on the results. The study finds that a Kolmogorov-like slope
for the longitudinal velocity spectrum can be reached for a limited
range of wave numbers, but the solenoidal power spectrum appears to be
rather strongly affected by numerical resolution effects (in fact, the
measured slopes are much steeper than Kolmogorov for all viscosity
schemes, and this was found to be robust as a function of resolution
over the tested range).  Overall, \citet{Valdarnini2011} finds that
turbulent motions account for a few to 10\% of the thermal energy
content.
 
Mesh-based studies of cluster turbulence find similar or even higher
contributions of the kinetic energy in turbulence relative to the
thermal energy in the ICM \citep{Iapichino2008,
  VazzaBrunetti2009,Maier2009, Lau2009, VazzaBrunetti2011, Paul2011,
  Schmidt2011, Iapichino2011, Jones2011}.  For example,
\citet{VazzaBrunetti2011} report 5 to 30\% based on simulations with
the {\small ENZO} code.  All these studies agree that major mergers
efficiently inject turbulence, and produce a radial trend where the
importance of turbulence increases with radial distance.  In
particular, \citet{Paul2011} point out that there is a close
connection between turbulence production and virialization.  Also,
they show that the turbulence after a major merger is quite
long-lived, still accounting for about 15\% of the thermal energy
after $4\,{\rm Gyrs}$ and 5\% after $10\,{\rm Gyrs}$.  \citet{Zhu2010,
  Zhu2011} have examined turbulence and vorticity in the intergalactic
medium, claiming that at $z=0$ the IGM is in a fully turbulent state
on scales less than about $\sim 3\,{\rm Mpc}$ and that this
significantly modifies structure formation and the gas fractions in
low mass halos.

\begin{table}
\begin{tabular}{lcccc}
\hline
$\mathcal{M} \sim \mach$ turbulence simulations\\ 
\hline
       & \multicolumn{4}{c}{resolution elements} \\ 
      & $64^3$  & $128^3$  & $256^3$  & $512^3$  \\ 
\hline
SPH  &   {S1} &   {S2} &   {S3} &   {S4}   \\
SPH (time dependent AV)  &   {S1-tav} &   {S2-tav} &   {S3-tav}     \\
{\small AREPO} (moving-mesh)  &   {A1} &   {A2} &   {A3} &   {A4}   \\
Fixed Cartesian grid  &   {F1} &   {F2} &   {F3} &   {F4}   \\
\hline
\end{tabular}
\caption{Names of primary simulation runs for $\mathcal{M} \sim \mach$
  turbulence. We consider calculations with three different numerical
  methods, (1) SPH with default parameters as implemented in the
  {\small GADGET} code and with a time dependent artificial
    viscosity parametrization, (2) the moving-mesh approach of
  {\small AREPO}, and (3) a fixed Cartesian mesh, which is also
  realized with {\small AREPO}.}
\label{tab:subsim}
\end{table}

\begin{table}
\begin{tabular}{lll}
\hline
\multicolumn{3}{l}{$\mathcal{M} \sim \mach$ turbulence, variants of SPH simulations}\\ 
\hline
Run name  & resolution & Characteristics  \\
\hline
S2-ngb1 &$128^3$& $N_{\rm ngb}=180$ smoothing neighbours\\
S2-ngb2 &$128^3$&  $N_{\rm ngb}=512$ smoothing neighbours\\
S3-$\alpha = 0.1$ &$256^3$&  reduced viscosity coefficient $\alpha = 0.1$\\
S3-$\alpha = 0.01$ &$256^3$&  reduced viscosity coefficient $\alpha = 0.01$\\
S3-$\alpha = 0.001$ &$256^3$&  reduced viscosity coefficient $\alpha = 0.001$\\
S3-$\alpha = 0.0001$ &$256^3$&  reduced viscosity coefficient $\alpha = 0.0001$\\
S3-balsara &$256^3$&  enabled Balsara shear viscosity factor\\
S3-tav-balsara &$256^3$&TAV + Balsara shear viscosity factor\\
\hline
\end{tabular}
\caption{Variations of the numerical parameters of our standard SPH
  simulation for driven $\mathcal{M} \sim \mach$ turbulence. For
  resolutions of  $128^3$ or $256^3$, we carry out several simulations where either the number of
  SPH smoothing neighbours, or the artificial viscosity parameterization
  is changed, as indicated in the table.}
\label{tab:sphsim}
\end{table}

\begin{table}
\begin{tabular}{lccc}
\hline
Mach number          &   $\mathcal{M} \sim \mtrans$   &  $\mathcal{M} \sim \mmed$ & $\mathcal{M} \sim \mhigh$  \\ 
\hline
SPH                  & S3-m1                    &  S3-m5              & S3-m10  \\
{\small AREPO} (moving-mesh)  & A3-m1                    &  A3-m5              & A3-m10  \\
Fixed Cartesian grid & F3-m1                    &  F3-m5              & F3-m10  \\
\hline
\end{tabular}
\caption{Simulations carried out for transsonic and supersonic driven
  turbulence. For the three Mach numbers examined here,
  $\mathcal{M} \sim \mtrans$, $\mathcal{M} \sim \mmed$ and
  $\mathcal{M} \sim \mhigh$, we carry out simulations at a fixed
  nominal resolution of $256^3$ particles/cells with three different methods,
  SPH, a moving mesh, and a fixed Cartesian mesh.
  \label{tab:supersim}}
\end{table}

\begin{table}
\begin{tabular}{llccc}
\hline
& &         $\mathcal{M} \sim \mach $ & $\mathcal{M} \sim \mtrans - \mmed$ &  $\mathcal{M} \sim \mhigh$\\
\hline
& $\sigma$   & $0.014$   & $0.21$ / $3.0$ & $12.247$           \\
& $\Delta t$ & $0.005$   & $0.005$   & $0.005$\\
& $t_s$      & $1$       &  $0.5$    & $0.05$\\
& $k_{\rm min}$ & $6.27$  & $6.27$ & $6.27$\\
& $k_{\rm max}$ & $12.57$ & $12.57$ & $18.85$\\
& $k \propto$ & $k^{-5/3}$ & $k^{-5/3}$ & $- (k-k_{\rm c})^2$ \\
\hline
\end{tabular}
\caption{Summary of the parameters of the turbulent driving routine used in our simulations.}
\label{tab:driving}
\end{table}

Unlike in the case of supersonic turbulence, there is a paucity of
systematic examinations of the accuracy of different numerical
techniques for representing subsonic turbulence. Only a few studies
have considered subsonic turbulence in SPH thus far
\citep{Violeau2007,Monaghan2011,Robinson2011} and we are not aware of
a comparative analysis of three-dimensional simulations in this
regime.  However, an examination of this question is particularly
timely because serious differences between Eulerian mesh-codes and
SPH have been reported in a number of recent papers
\citep[e.g.][]{Agertz2007}.  Given also that numerical differences in
the representation of turbulence have been suspected to significantly
influence cooling rates of halos and therefore impact galaxy formation
\citep{Vogelsberger2011, Sijacki2011, Keres2011}, it is important to
address this gap in detail.

In this study, we therefore compare the behaviour of turbulence
simulations with three different numerical methods. We use the
smoothed particle hydrodynamics implementation of the widely employed
{\small GADGET} code and compare it with the novel {\small AREPO}
technique \citep{Springel2010}, both using a moving mesh or a fixed
Cartesian mesh.  We primarily focus on the poorly studied subsonic
regime, but we also perform some simulations of transsonic and
supersonic turbulence to be able to compare our results with the
recent literature, and to characterize the behaviour of our new
{\small AREPO} code in this regime as well. In this paper, we will
only examine the well established ``standard formulation'' of SPH
\citep[as described, for example, in][]{Springel2002}, supplemented
also with a time dependent artificial viscosity parameterization. We
note however that a number of recent works proposed extensions or
modifications of classic SPH that aim to improve the accuracy of this
method \citep[e.g.][]{Wadsley2008,Price2008, Hess2010, Read2010,
  Read2011, Abel2011}. Our results do not necessarily extend to these
new flavours of SPH, and it remains to be seen weather they can
resolve the problems pointed out here.

This paper is organized as follows. In Section~\ref{sec:meth}, we
outline our numerical techniques and describe our simulation set, as
well as our analysis techniques.  In Section~\ref{sec:sub}, we present
our results for simulations in the subsonic regime. We then turn to
results for the transsonic and supersonic regimes in
Section~\ref{sec:super}. Finally, we give a discussion of our findings
and present our conclusions in Section~\ref{sec:disc}.
 
\section{Methodology} \label{sec:meth}
 
\subsection{Numerical methods}

\subsubsection{Moving- and fixed-mesh simulations with {\small AREPO}}

{\small} AREPO implements a novel quasi-Lagrangian scheme for solving
the Euler equations on an unstructured moving mesh, as described in
detail in \citet{Springel2010}.  The mesh is defined as the Voronoi
tessellation of a finite set of points that are distributed in the
simulation volume. A finite volume approach for hydrodynamics is
formulated on this mesh, based on a Godunov-type scheme with
second-order accurate reconstruction and an exact Riemann solver applied to all
interfaces for estimating hydrodynamical fluxes. 

If the mesh-generating points are kept stationary, this hydrodynamical
solver is equivalent to the MUSCL-Hancock second-order accurate scheme
widely employed in many Eulerian hydrodynamics codes on Cartesian
meshes. In fact, this equivalence can be made exact if the
mesh-generating points are arranged on a Cartesian grid, in which case
the cells of {\small AREPO} also become Cartesian. We will carry out
some of our simulations in this mode, which we shall refer to as
``fixed-mesh''. However, the novel aspect of {\small AREPO} is that
the mesh-generating points are allowed to move freely, without
producing problematic mesh-twisting effects. In particular, the points
can be moved with the local fluid velocity itself, thereby producing
an adaptive, quasi-Lagrangian behaviour. In this ``moving-mesh'' mode,
advection errors are greatly reduced and become in fact independent
of the presence of a possible bulk flow, making the results of {\small
  AREPO} manifestly Galilean-invariant, unlike ordinary Eulerian
codes.

{\small AREPO} can additionally employ on-the-fly refinement and
derefinement operations of its mesh, similar to adaptive mesh
refinement (AMR) methods. We invoke this in our moving-mesh simulations
to guarantee that the mass resolution is always approximately
constant, as in the SPH simulations that we compare with. To this end,
cells are (de)refined if their mass deviates by more than a factor of
two from the desired target mass resolution (which is the initial cell
mass). We note however that such (de)refinement operations are only
rarely needed because the Lagrangian mesh motion already yields a nearly
constant mass per cell. We also make use of {\small AREPO}'s mesh
regularization feature, where mesh-generating points of highly
distorted cells may receive an additional small velocity component
towards the geometric center of their cell. This results in a more
regular mesh, which reduces errors in the linear reconstruction step.

We note that the {\small AREPO} code has recently been successfully
used in first science applications, studying first star formation
\citep{Greif2011} and galaxy formation
\citep{Vogelsberger2011}. There also already exist extensions to
include magnetohydrodynamics \citep{Pakmor2011}, radiative transfer
\citep{Petkova2011}, as well as treatment of the full Navier-Stokes
equations \citep{Munoz2011}.

\subsubsection{Smoothed particle hydrodynamics}

Smoothed particle hydrodynamics (SPH) is a particle-based approach to
fluid dynamics which is popular in astronomy due to its geometric
flexibility, automatic adaptivity, and good conservation properties
\citep[see e.g.][for recent reviews]{Rosswog2009, Springel2010b}. We
use the simulation code {\small GADGET-3} \citep[last described
in][]{Springel2005} for our SPH simulations, which employs a
``standard'' formulation of SPH with fully adaptive smoothing lengths
and a simultaneous conservation of entropy and energy
\citep{Springel2002}.  We note however that there are many
  alternative formulations of SPH, some of them quite recently
  proposed to address certain accuracy problems of SPH
  \citep{Price2008, Read2010, Read2011, Hess2010}.  Our
  results may not necessarily apply to all of these flavours.

In some of our simulations, we also study the influence of numerical
parameters in SPH on our results, such as the number
$N_{\textrm{sph}}$ of smoothing neighbours and the artificial
viscosity parameterization. In {\small GADGET-3}, the SPH smoothing
lengths $h_i$ of particles are adjusted such that $(4\pi/3) h_i^3
\rho_i = N_{\textrm{sph}}\overline{m}$ is always fulfilled, where
$h_i$ is the radius at which the smoothing kernel drops to zero,
$\rho_i$ is the density estimate of the particle $i$, and
$\overline{m}$ is the target mass resolution (here equal to the SPH
particle masses). In our default 3D simulations we use
$N_{\textrm{sph}} = 64$ smoothing neighbours.

The artificial viscosity is implemented as a viscous force:
\begin{equation}
  \left.\frac{\dd \vect{v}_i}{\dd t} \right|_{\textrm{visc}} = - \sum_j m_j
  \Pi_{ij} \nabla_i \overline{W}_{ij} ,
\end{equation}
where $\overline{W}_{ij}$ is the arithmetic average of the smoothing
kernels and $\Pi_{ij}$ parameterizes the viscous tensor.  We use the
following form \citep{Monaghan1997, Springel2005} for $\Pi_{ij}$ in our default
runs:
\begin{equation}
  \Pi_{ij} = - \frac{\alpha}{2} \frac{(c_i+c_j-3w_{ij})\cdot w_{ij}}{\rho_{ij}},
\end{equation}
with $w_{ij} = \vect{v}_{ij}\cdot \vect{r}_{ij}/|\vect{r}_{ij}|$ if
$\vect{v}_{ij}\cdot \vect{r}_{ij} < 0$, otherwise $w_{ij}=0$. For this
definition of $w_{ij}$, the artificial viscosity is always repulsive,
and is non-zero only if a pair of particles approaches each other, implying
that the entropy produced by the viscosity is positive definite.

One general problem of artificial viscosity parameterizations is that
they may introduce spurious viscosity also outside of shocks, in
regions where it should in principle not be needed
\citep[e.g.][]{Cullen2010}. This can be a significant problem in shear
flows, where this effect may lead to unwanted angular momentum
transport. To suppress the artificial viscosity in regions of strong
shear, \cite{Balsara1995} proposed a simple viscosity limiter in the
form of an additional multiplicative factor $(f_i+f_j)/2$ for the
viscous tensor, defined as
\begin{equation}
  f_i = \frac{|\nabla \cdot \vect{v}|_i}{|\nabla \cdot \vect{v}|_i+|\nabla \times \vect{v}|_i}.
\end{equation}
This limiter is often used in cosmological SPH simulations and also
available in the {\small GADGET} code. In our default simulations with
fixed $\alpha$, we have refrained from enabling it, but we have also
run comparison simulations where it is used, as discussed in our
results section.

In addition, we consider a so-called time variable artificial
  viscosity, as first proposed by \citet{Morris1997}. The idea is here
  to try to reduce the viscosity in regions away from shocks such that
  it is applied in a more targeted fashion only where it is really
  needed. We employ the implementation of \citet{Dolag2005} in the
  {\small GADGET} code, where $\alpha$ is replaced with an individual
  parameter $\alpha_i(t)$ for each particle:
\begin{equation}
\frac{{\rm d} \alpha_i}{{\rm d}t} = -\frac{\alpha_i-\alpha_{\rm
    min}}{\tau}+S_i .
\end{equation}
 The time evolution is controlled by a source term $S_i$ that ramps up
 the viscosity quickly when a fast compression is detected, and a
 decay function that makes the viscosity decline again in smooth
 regions of the flow to a small minimum value $\alpha_{\rm min}$ over
 the decay time $\tau$. We note that a near-identical formulation
 has also been used by \citet{Price2012}.  

\subsection{Turbulent driving}

In this work, we consider isothermal gas in which turbulence is
induced through an external stochastic forcing on large scales. The
condition of isothermality is not crucial for our study of subsonic
turbulence, but it conveniently prevents that the turbulent kinetic
energy dissipated in the flow leads to a gradual increase of the
pressure in the gas with time. Instead, the dissipated energy is
simply lost from the isothermal system, so that a statistically 
quasi-stationary state of developed turbulence can be reached after
some time, where on average the energy injected on large scales is
lost  on smaller scales by dissipation.

Our method for calculating the acceleration field follows closely the
procedure used in
\citet{Schmidt2006,Federrath2008a,Federrath2009,Federrath2010} and
\citet{Price2010}.  In particular, the acceleration field is setup in
Fourier space and only contains power in a small range of low
frequency modes between $k_{\rm min}=6.27$ and $k_{\rm
  max}=12.57$. The relative amplitude of the forcing modes over this
small range is varied as $P(k) \propto k^{-5/3}$. Except in our run at
$\mathcal{M} \sim \mhigh$, $P(k)$ is a paraboloid centred around
$(k_{\rm min}+k_{\rm max})/2$ with $k_{\rm min}=6.27$ and $k_{\rm
  max}=18.85$.  The phases of the Fourier modes are drawn from an
Ornstein--Uhlenbeck process and are periodically updated after a time
interval $\Delta t$. The corresponding random sequence is given by
\begin{equation}
\vect{x}_{t}=f\, \vect{x}_{t-\Delta t} + \sigma \sqrt{(1-f^2)}\, \vect{z}_n ,
\end{equation}
where $f$ is a decay factor given by $f=\exp(-\Delta t/t_s)$, with
$t_s$ being the correlation length.  $\vect{z}_n$ is a Gaussian random
variable and $\sigma$ is the variance of the Ornstein--Uhlenbeck process.
The resulting sequences have zero mean, $\left <\vect{x}_t\right >=0$,
and their correlations are given by
$\left<\vect{x}_t\,\vect{x}_{t+\Delta t}\right> = \sigma^2 f$. The
frequent but correlated changes of the acceleration field as a
function of time result in a smoothly varying turbulent driving field.

We use a purely solenoidal driving in this study, which can be
obtained by projecting out the compressive part of the acceleration
field through a Helmholtz decomposition in $k$-space.  The projection
operator is given by
\begin{equation}
  \vect{\hat a}(\vect{k})_i =\left(\delta_{ij} - \frac{k_ik_j}{|{k}|^2}\right) \vect{\hat a_0}(\vect{k})_j
\end{equation}
in Fourier space. We note that solenoidal driving appears particularly
appropriate for subsonic turbulence. Compressive modes would only
cause additional sound waves and only start to couple to smaller modes
once non-linear steepening of these acoustic waves becomes important.
In any case, if a compressive component was added, we would expect a
somewhat broader density PDF for a given Mach number
\citep{Federrath2008a}.

Finally, the acceleration field due to the driving mechanism is
calculated in position space at each particle or cell position
directly as a sum over the small number of non-zero Fourier modes, a
procedure which is free of any resolution limitations.  This field is
then introduced as an additional source term in the Euler equations,
in the same way as an external gravitational field is normally coupled
to gas dynamics. In our time-integration scheme, the external
acceleration is added in two half-steps at the beginning and end of
each timestep, producing a leap-frog type integration scheme.

\subsection{Simulation set}

We consider periodic boxes of unit length on a side, filled with
isothermal gas at unit mean density and unit sound speed
($\overline{\rho} = c_s = 1$). The Euler equations are
scale-independent, so that once quasi-stationary turbulence has
developed the only characteristic parameter remaining is the mean Mach
number $\mathcal M$, which we define as the mass-weighted rms-velocity
relative to the sound speed. The amplitude of the driving field
determines $\mathcal M$, and can in principle be freely adjusted to
reach the desired strength of turbulence.  We note that we have here not attempted
to subtract the mean gas velocity in our simulations even though a
respectable amount of bulk motion can be generated in the supersonic
regime as a result of our driving.  In fact, the kinetic energy of the
bulk motion can become at times nearly as large as the kinetic energy
of the irreducible smaller scale motions, and hence matters when
measuring the mean Mach number in terms of the total kinetic energy. 
The bulk motion is negligible in the subsonic regime.
As the bulk motion however only affects the DC mode in Fourier space,
it has no influence on our power spectrum measurements.  The Mach
numbers we report are corrected for the bulk motion and do not include
it.

For technical reasons having to do with our measurement technique for
dissipation discussed below, we actually do not use an exact
isothermal equation of state, but rather one with an adiabatic index
of $\gamma=1.001$, combined with enforcing the entropy of the gas to
stay at the initial value after completion of each
timestep. Specifically, for a prescribed initial mean density
$\overline{\rho}$, and an (isothermal) sound speed $c_s$, we
initialize the gas with a specific entropy
\begin{equation}
\overline{A} = c_s^2 \overline{\rho}^{1-\gamma}, 
\label{eqn:entropy}
\end{equation}
hence the pressure is  given by
\begin{equation}
P_i = \overline{\rho} c_s^2 \left(\frac{\rho_i}{\overline{\rho}}\right)^\gamma 
\label{eqn:pressure}
\end{equation}
for a cell/particle of density $\rho_i$. The specific internal energy
per unit mass, $u_i$, can be calculated from the specific entropy as
\begin{equation}
u_i = \overline{A} \frac{\rho_i^{\gamma-1}}{(\gamma-1)}.
\end{equation}
In our `quasi-isothermal' simulations, the entropy of the gas is
reset to $\overline{A}$ after each timestep, so that the pressure of a
particle/cell is always given by equation~(\ref{eqn:pressure}).  All
our simulations are started from a regular Cartesian grid of particles
with initially zero velocities, and use a global time step for all
particles/cells.

Each of our primary subsonic simulations was performed with three
different numerical methods: SPH as implemented in the {\small
  GADGET-3} code, moving-mesh hydrodynamics using {\small AREPO}, or
fixed-mesh Eulerian hydrodynamics also based on {\small AREPO} but
with a stationary Cartesian mesh. The resolution of our simulations
ranges form $64^3$ to $512^3$ particles or cells, respectively.
Table~\ref{tab:subsim} gives an overview of these simulations, and
lists their most important parameters. Our naming convention is such
that a leading capital letter indicates the type of a simulation (`S'
for SPH, `A' for moving-mesh with {\small AREPO}, and `F' for a fixed
mesh), followed by a digit that indicates the resolution level (`1'
for $64^3$, `2' for $128^3$, etc.). When we compare different
simulation techniques or different numerical resolutions, we always do
this at identical driving amplitude, such that any difference that is
seen arises from the hydrodynamics alone. In particular, all of our
subsonic simulations (where $\mathcal{M}\sim \mach$) use an identical
turbulent forcing field.

In our default SPH simulations, we consider two different artificial
viscosity formulations, one with a viscosity strength parameter equal
to $\alpha=1$, the other with a time-variable viscosity that is
individual for each particle (both with and without a shear viscosity
limiter). These choices approximately bracket the range of viscosity
settings that are in use in production calculations in cosmology.  In
order to examine the dependence of our results on numerical parameters
of SPH in more detail, we have additionally performed a set of SPH
simulations with further variations in the artificial viscosity
parameters ($\alpha=1.0$ with a shear viscosity limiter, $\alpha=0.1$,
$\alpha=0.01$, $\alpha=0.001$, and $\alpha=0.0001$). Furthermore, we
have also studied the influence of the number of SPH smoothing
neighbours. Our default value for $N_{\mathrm{ngb}}$ is $64$, but we
also used $N_{\mathrm{ngb}} = 180$ and $N_{\mathrm{ngb}} = 512$. The
corresponding simulation runs and their symbolic names are summarized
in Table~\ref{tab:sphsim}.

 \subsection{Measuring dissipation}
\label{sec:meth:diss}
 The classical theory of Kolmogorov for incompressible turbulence
 conjectures that energy is injected on large scales and then cascades
 down to eddies of ever small size, until dissipation on very small
 scales eventually occurs.  In this picture, large scale gas motions
 in the resulting turbulent cascade do not dissipate energy in any
 significant way, instead the energy is essentially transported in a
 conservative fashion over the inertial range down to the dissipation
 scale.  In the analysis of numerical turbulence simulations it is
 standard procedure to examine how the kinetic energy is distributed
 as a function of scale, which is usually done in terms of the
 velocity power spectrum. We suggest here that it is also interesting to
 try to directly measure the energy dissipation as a function of
 scale, as this provides interesting complementary information about
 the dissipative properties of a numerical scheme.

 To this end, we define dissipation as the irreversible conversion of
 kinetic energy into heat.  Because we use $\gamma = 1.001$ in our
 ideal gas equation-of-state, dissipation manifests itself as an
 increase in the specific entropy of a cell or a particle. To measure
 this quantity, we compare $A_i$ after completion of every timestep
 with the value $\overline{A}$, afterwards resetting $A_i$ to
 $\overline{A}$. The extracted thermal energy, $\Delta E_i = m_i
 (A_i-\overline{A}) \rho_i^{\gamma-1} / (\gamma-1)$, is then the {\em
   dissipated} energy, which we assume to leave the system in
 concordance with the quasi-isothermal conditions we impose.

 In SPH, $A_i$ only changes due to the artificial viscosity and is
 easily obtained from the work done against the artificial viscosity
 forces. The $\Delta E_i$ measured for a SPH particle is always
 positive definite in this case. In our mesh-code {\small AREPO}, exact
 energy-, mass- and momentum-conservation is given in every
 hydrodynamic step. To measure the dissipative increase of entropy of
 a cell, we additionally advect the entropy in the system in a
 conservation fashion, as described by \citet{Springel2010}. The
 dissipative energy change $\Delta E_i$ of a cell can then be
 estimated in the above fashion, with the caveat that the energy
 $\Delta E_i$ is not guaranteed to be positive definite for all cells
 due to discretization errors.  However, in this case a local average
 over a group of cells will still give a faithful estimate of the
 total energy that is lost, due to the manifestly conservative
 properties of the mesh-based evolution of the fluid. We shall use the
 $\Delta E_i$ values for a Fourier analysis of the spatial scales on
 which most dissipation occurs, and for cross-checking whether the
 total dissipated energy balances the total injected energy when
 steady-state turbulence is reached.

\subsection{Power spectrum measurements}
\label{sec:meth:power}

\begin{figure}
\begin{center}
\setlength{\unitlength}{1cm}
\includegraphics{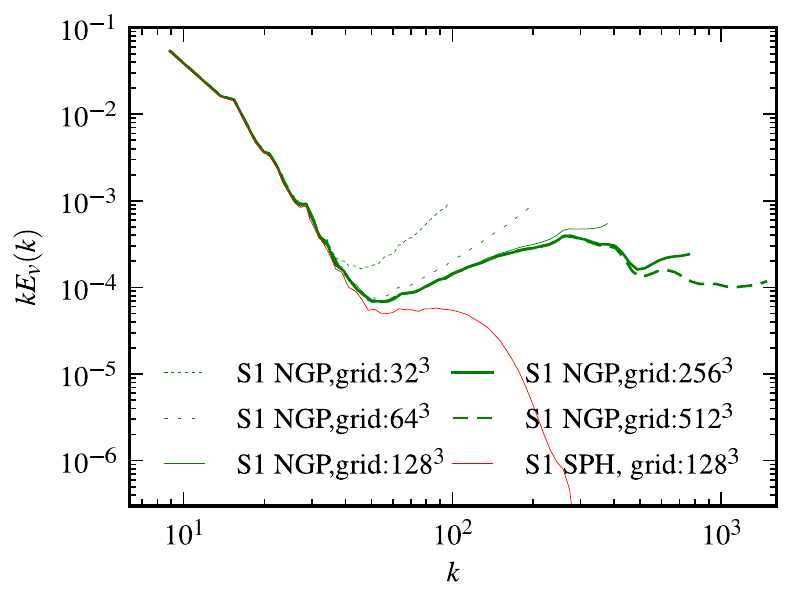}
\caption{
Results for different approaches to measure the velocity
  power spectrum for one of our $64^3$ SPH simulations. The green
  lines with different line styles show our nearest point sampling,
  with sampling resolutions $32^3$, $64^3$, $128^3$, $256^4$ and $512^3$. The
  results for $128^3$ (our default grid size at this resolution) and higher are virtually identical up to the
  Nyquist frequency of the run. The red line shows a measurement when
  the velocity field is calculated by SPH-smoothing for a $128^3$
  mesh. This method \citep[advocated by][]{Price2012}
  suppresses small-scale velocity noise but also removes kinetic
  energy associated with particle motions on these scales.}
\label{fig:meth:power}
\end{center}
\end{figure}

The 3D power spectrum of a scalar or vector field $w$ is defined as
the Fourier transform of the two-point correlation function
\begin{equation}
C_w(\vect{l}) = \langle  w(\vect{x}+\vect{l})w(\vect{x}) \rangle_{\vect{x}}.
\end{equation}
Thus
\begin{align}
E_w(\vect{k}) &= \frac{1}{(2 \pi)^3} \int_V C_w(\vect{l}) \exp (-i
\vect{k} \vect{l})\, {\rm d}^3\vect{l}\\
                &= \left( \frac{2 \pi}{L}\right)^3 \left |\hat{w}({\vect k}) \right |^2 ,
\end{align}
where $\hat{w}$ is the Fourier transform of $w$.  In order to
numerically estimate $\hat{w}$, the field $w$ is usually represented in
discretized form on a Cartesian grid, allowing an efficient
measurement of the power spectrum through discrete Fourier transforms.

We here use a nearest neighbour sampling of the intrinsic
hydrodynamical quantities of the particle/cell data of our simulations
to do this. The value of the desired quantity $w$ at each grid cell is
hence obtained as the value of the particle or cell closest to the
centre of a grid cell.  This simple approach aims to maximize the
information content extracted from the simulations, but risks to
suffer from power aliasing effects if the employed Fourier grid
is too small. We have however checked that our default Fourier mesh
size we employ for our power spectrum measurements (based on a grid
twice as fine as the resolution of the simulation at hand) is fine
enough to make such effects negligible. 

 Another important aspect of our power spectrum measurement is that it
 faithfully represents the total kinetic energy of the particle/cell
 set. This should be given as the integral over the measured power
 spectrum. According to Parseval's theorem, this integral is equal to
 the variance of the velocity field that is used to measure the
 spectrum through a discrete Fourier transform. As our velocity field
 definition creates a fair sample of the particle/cell velocity
 values, we obtain an accurate accounting of the total kinetic energy
 in the flow.  In contrast, smoothing the SPH velocity field via
 kernel interpolation as advocated by \citet{Price2012} removes
 kinetic energy on small scales and can cause a significant error in
 the total energy represented by the power spectrum measurement. This
 is explicitly demonstrated in Figure \ref{fig:meth:power}, where we
 show a power spectrum measurement for one of our $64^3$ SPH
 simulations carried out with different techniques. The green lines
 with different line styles show our nearest point sampling, with
 resolutions $32^3$, $64^3$, $128^3$, $256^4$ and $512^3$. The results
 for $128^3$ and higher are virtually identical up to the Nyquist
 frequency of the run and accurately reflect the total kinetic energy
 of the particle set. On the other hand, if we SPH-smooth the velocity
 field on a $128^3$ mesh, we obtain the red solid line. While this
 method suppresses small-scale velocity noise, it also reduces the
 kinetic energy in the field by $\sim 3.6\%$ in this case.  However,
 if for example our S1-tav simulation is considered the suppressed
 energy amounts to about $\sim 13\%$ of the total kinetic energy due
 to the higher power on smaller scales in that simulation.  We note
 that we calculate our power spectrum measurements on-the-fly while
 the turbulence simulations are run, allowing us to reach a very fine
 temporal resolution for the evolution of the power spectrum of the
 different quantities we examine.

Assuming an isotropic statistical distribution, the 1D power spectrum
of the quantity $w$ can then be obtained by angular averaging
$E_w(\vect{k})$ over shells in $k$-space, as
\begin{equation}
E_w(k) = 4 \pi k^2 \langle E_w({\vect k}) \rangle
\end{equation}
where $k = \left | {\vect k} \right |$.  We employ fine logarithmic bins in
$k$ for determining the mean power per mode at a certain $k$, with
bins combined as needed such that a minimum number of modes per
$k$-bin is obtained. 
The normalization of the 1D power spectrum is chosen such that the
integral over the power spectrum gives always the total power, i.e.
\begin{equation}
  \sigma^2 = \int E_w(k) \, {\rm d}k = \frac{1}{N^3}
  \sum_{j,k,l=0}^{N-1} |w_{ijk}|^2  .
\end{equation}

We note that we usually plot the quantity $k E_w(k)$ in our power
spectrum plots as a function of $\log k$, instead of using $E_w(k)$
directly. This has the advantage that a horizontal line in such a plot
corresponds to constant total power per logarithmic decade, one with a
positive slope means that small scales dominate, whereas a negative
slope indicates that the total power in the examined quantity is
dominated by large scales.

\begin{figure}
\begin{center}
\setlength{\unitlength}{1cm}
\includegraphics{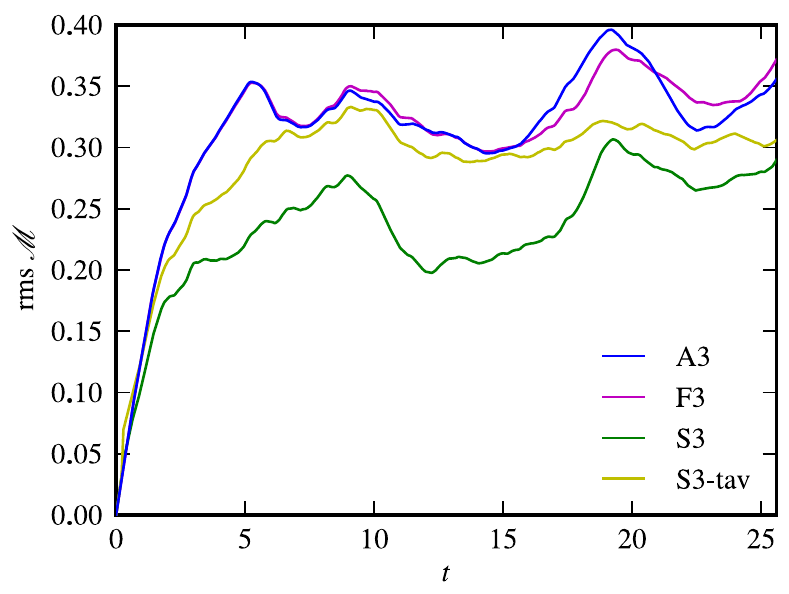}
\caption{Mean Mach-number evolution as a function of time for our
  subsonic turbulence simulations. Here $\mathcal{M}$ is defined as
  the mass-weighted rms velocity in units of the sound speed. The
  different lines give results for SPH (green and yellow), {\small
    AREPO} (blue), and a fixed-mesh (magenta), at a resolution of
  $256^3$ particles/cells (runs at different resolutions give
  extremely similar results). We see that a quasi-stationary state is
  reached after time $t\sim 10$, but the total kinetic energy in the
  $\alpha=1.0$ SPH case (green) is somewhat smaller than in the two
  mesh codes  and in the SPH run with time-variable artificial
    (tav) viscosity (yellow). }
\label{fig:sub:mach} 
\end{center}
\end{figure}

\begin{figure}
\begin{center}
\setlength{\unitlength}{1cm}
\includegraphics{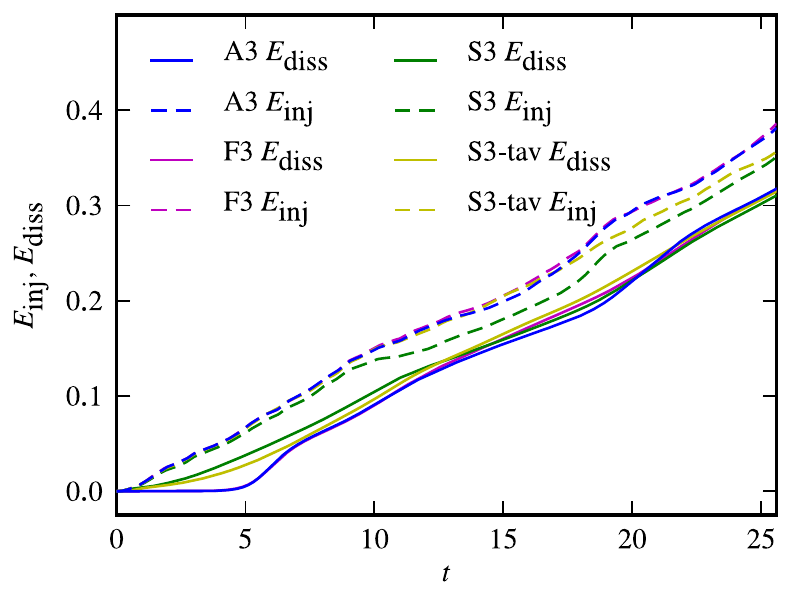}
\caption{Time evolution of the total cumulative injected energy
  (dashed lines) and the total dissipated energy (solid lines), for
  different simulation runs, as labeled. At any given point in time,
  the difference between the injected energy and the dissipated energy
  is the kinetic energy stored in the simulation box. The time
    average energy dissipation rate per unit mass is $\epsilon \simeq 0.016$. }
\label{fig:sub:dis}
\end{center}
\end{figure}

\begin{figure*}
\begin{center}
\setlength{\unitlength}{1cm}
\includegraphics{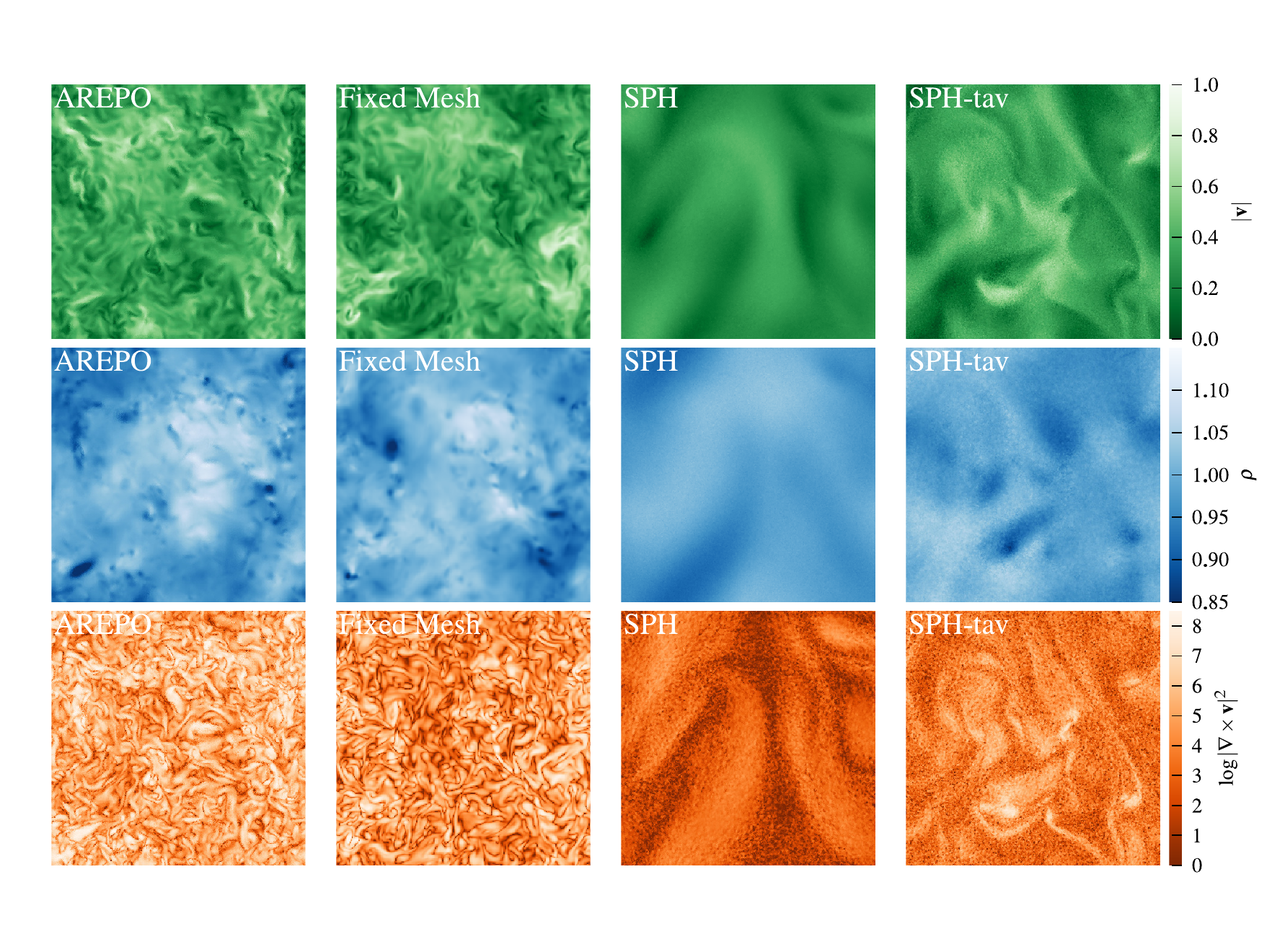}
\caption{Visual comparison of the turbulent velocity field (top row),
  the density field (middle row) and the enstrophy $|\nabla \times
  \vect{v}|^2$ (bottom row) in quasi-stationary turbulence with
  $\mathcal{M} \sim \mach$, simulated with different numerical
  techniques. Shown are thin slices through the middle of the
  periodic simulation box at the final time $t=25.6$. From left to right, we show our moving
  grid result  (A3), an equivalent calculation on a static mesh (F3), and  two SPH
  calculations, one with a fixed $\alpha=1.0$ viscosity (S3) and the
    other with time-variable viscosity  (S3-tav), as labeled.}
\label{fig:sub:velocity}
\end{center}
\end{figure*}

Finally, we want to clarify how we measure the power spectrum of the
energy dissipation rate, which requires a special treatment. In order
to allow for a direct comparison with the kinetic energy power
spectrum obtained from $w = \vec{v}$, we in principle want to set $w =
\sqrt{\Delta E}$, where $\Delta E$ is the energy dissipation rate
measured as described above. However, since the measured dissipation
rate can sometimes exhibit negative values in the case of the mesh
code, this procedure needs to be modified. We thus compute the power
spectrum for $w_+ = \sqrt{\Delta E_+}$ and $w_- = \sqrt{\Delta E_-}$
separately, where $\Delta E_{+}=\mathrm{min}(\Delta E,0)$ and $\Delta
E_{-}=\mathrm{max}(\Delta E,0)$ are the positive and negative parts of
the measured dissipation field.  Finally, the dissipation power
spectrum is then estimated as
\begin{equation}
E_{\mathrm{diss}}(k) = E_{w_+}(k)-E_{w_-}(k).
\end{equation}
Note that the $k$-integral of this quantity is equal to the total
energy dissipation rate.

\section{Subsonic turbulence}  \label{sec:sub}

\subsection{Global characteristics}

In Figure~\ref{fig:sub:mach}, we show the time evolution of the rms
Mach number for our runs of subsonic turbulence at a resolution of
$256^3$ (S3, S3-tav, A3, and F3). After an initial ramp up of the turbulent
energy, a quasi-stationary state is established, starting at time
$t\sim 5-10$. There are however still substantial intermittent
fluctuations in the global rms Mach number, making it clear that
averaging over extended periods of time is required to obtain truly
stable results for the statistical properties of the turbulent fluid
state, especially on large scales.  We note that runs carried out with
different numerical resolutions give extremely similar results to the
ones shown in Fig.~\ref{fig:sub:mach}.  Interestingly, the time
evolutions of the moving-mesh and the fixed-mesh results agree very
well with each other, but the terminal Mach number reached by SPH is
significantly lower,  especially in the run with $\alpha = 1.0$.
This is despite the fact that the driving field imposes exactly the
same accelerations in all the simulations. The smaller overall kinetic
energy achieved in the SPH run with $\alpha=1.0$ is a result of viscous damping of
large-scales modes at or close to the driving scale. This effect
  is greatly reduced but not completely eliminated with the
  time-variable viscosity parameterization.

We show the cumulative injected and dissipated energy as a function of
time in Figure \ref{fig:sub:dis} for the same simulations. Note that
the difference between these two quantities is exactly the kinetic
energy stored in the gas at the corresponding time. Interestingly, the
mesh-based simulations do hardly dissipate any energy until $t=5$, in
contrast to the SPH simulations which show signs of energy dissipation
right from the start. This is consistent with the impression from
Figure~\ref{fig:sub:mach} that it is harder in SPH than in the
mesh-code to set the largest eddies into motion.  At around $t \sim
13$, the total cumulative dissipated energies begin to be rather
similar for all three methods, but the total injected energy of the
SPH simulations still lags behind the mesh-based runs.  This is simply
because the lower velocities in SPH reduce the average value of
$\left< \vect{v} \vect{a}_{\textrm{driv}} \right>$, where
$\vect{a}_{\textrm{driv}}$ is the acceleration due to the external
driving field.  As a result, the kinetic energy in the SPH runs never
fully manage to close the gap to the mesh-based calculations.

\subsection{Visualizations of the turbulent velocity, kinetic energy
  density and vorticity fields}

To better understand the systematic differences between the simulation
techniques, it is instructive to consider maps of fluid quantities in
slices through the simulation cube. To construct them, we use nearest
neighbour sampling at the coordinates of a two-dimensional grid with
$512^2$ pixels, twice finer than the nominal resolution of the
simulations examined here. Each pixel will hence show the value of the
closest mesh cell or SPH particle, respectively. This directly
reflects the individual fluid elements used in the discretization
schemes of the two numerical techniques, highlighting mesh or sampling
artefacts if they exist, as well as discretization noise. 

In the top row of Figure~\ref{fig:sub:velocity}, we show slices of the
velocity field at the final time $t=25.6$ of our subsonic turbulence
simulations, comparing the moving-mesh calculation with {\small AREPO}
to the one with a fixed Cartesian mesh, and to our SPH simulations
with $\alpha=1.0$ and time-variable viscosity. We can see that the
slices of the moving mesh (top left panel) and the fixed mesh
calculations (top middle panel) appear qualitatively very similar,
featuring both large-scale coherent motions and many irregular
small-scale features. The moving-mesh result appears slightly less
smooth and shows more small-scale features, but based on these images
alone it would be hard to decide whether this is due to a higher
effective resolution or due to the more irregular shaped cells in the
Voronoi case, which may induce some aliasing effects in the pixelized
map.

In contrast, the nearest neighbour SPH results for the velocity field,
shown in the two panels on the top right, look dramatically different. Here the
corresponding velocity fields do not contain the small-scale velocity
features present in the mesh-based calculations, particularly in the
$\alpha=1.0$ run, suggesting that an equally well developed
turbulent cascade has not really formed in the SPH simulation.

This impression is compounded by slices through the density and
enstrophy fields ($|\nabla \times \vect{v}|^2$), shown in the middle
and bottom row of Figure~\ref{fig:sub:velocity}, respectively. While
the mesh-based calculations exhibit a delicate mix of fine structures
in the kinetic energy density both on large and small scales, the SPH
$\alpha = 1.0$ simulation shows only some large-scale motions,
presumably reflecting primarily the driving field. While the S3-tav
run shows more structures, there is still a paucity of smaller flow
features.  Similarly, whereas the enstrophy slices reveal a granular
structure in the vorticity field that is dominated by small
structures, these are essentially completely absent in the SPH
calculations. The utilization of a time dependent artificial viscosity
scheme improves the SPH result noticeably. A further improvement might
be achieved through more advanced viscosity switches, like those
suggested by \citet{Cullen2010} or \citet{Read2011}.

\subsection{Velocity power spectra of subsonic turbulence}
\begin{figure}
\begin{center}
\setlength{\unitlength}{1cm}
\includegraphics{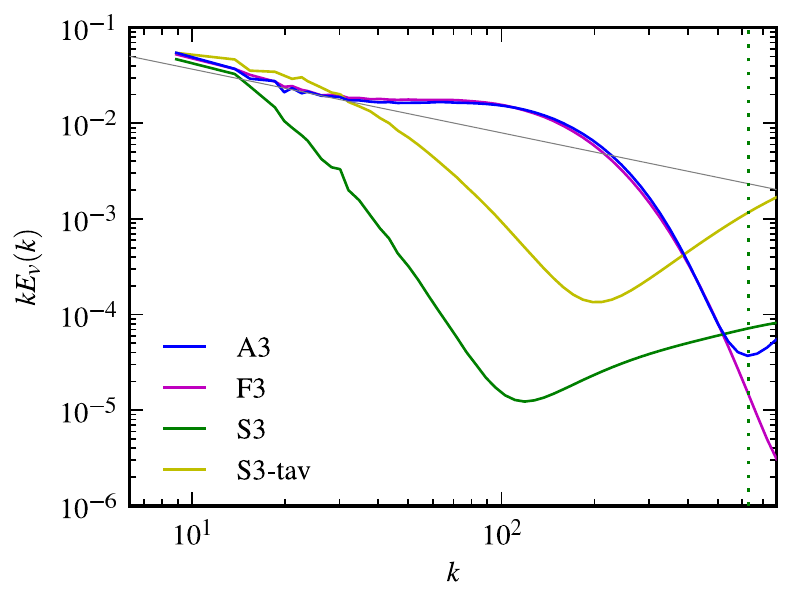}
\caption{Velocity power spectrum of SPH and {\small AREPO}, compared
  at a resolution of $256^3$.  The thin grey line shows the slope
  expected for Kolmogorov's theory of incompressible turbulence. The
  power in SPH falls much more rapidly than expected for fully
  developed turbulence. On small scales, the power rises again quite
  strongly up to the Nyquist frequency. This is small-scale velocity
  noise characteristic of SPH. The vertical green dotted line
  indicates the scale $2\pi/h_{\rm max}$, where $h_{\rm max}$ is the
  maximum SPH smoothing length among all the particles. }
\label{fig:sub:power256}
\end{center}
\end{figure}

A more quantitative analysis of this difference is obtained by
considering velocity power spectra of these four different simulation
techniques. In Figure~\ref{fig:sub:power256}, we compare power spectra
of the kinetic energy for our runs at $256^3$ resolution, averaged
over an extended period of time from $t=10$ to $t=25.6$, after a
quasi-stationary turbulent state has been established.

The results confirm the impression inferred from the previous
subsection.  There is a rather striking difference between the
mesh-based simulations and our SPH calculations.  The kinetic energy in
SPH is already drained at rather large scales, such that an extended energy
cascade is not formed. The self-similar turbulent power spectrum
expected based on Kolmogorov's theory for incompressible turbulence
($E_v(k) \propto k^{-5/3}$) is indicated as a thin grey power law in
the figure -- this line has a different slope compared with the
rapid decline of the power spectrum observed in SPH.

In contrast, the mesh-based simulations show a slope similar to the
expected Kolmogorov law, at least on very large scales.  Fits to the
power-law region of the velocity power spectrum give slopes of $-1.64$
and $-1.68$ for the moving mesh and the fixed mesh, respectively,
whereas SPH shows a wrong slope of $-4.14$ in the case of a constant
artificial viscosity and $-2.1$ in the case of a time-dependent
artificial viscosity.  There is even an excess of power in the
mesh-based results above the expected continuation of the Kolmogorov
slope, before the velocity power spectrum eventually starts to rapidly
decline on scales somewhat larger than the Nyquist frequency
corresponding to the nominal spatial resolution. This bump in power is
a manifestation of the so-called bottleneck effect, which is commonly
encountered close to the resolution limit in mesh-based studies of
turbulence and is also seen in experiments \citep[e.g.][and references
  therein]{Meyers2008}. The numerical bottleneck effect is similar to
the physical bottleneck effect and considerably complicates attempts
to robustly measure the true slope of the inertial range of
turbulence, as this requires the use of extremely high resolution
($2048^3$ and beyond), such that the bottleneck bump moves to
sufficiently small scales.  We note that the moving-mesh and
fixed-mesh calculations agree well with each other up to quite high
$k$, where eventually some small differences arise, caused by the
different mesh geometries and truncation errors in the two schemes.

Another interesting feature of the SPH velocity power spectrum is that
there is actually a minimum at some intermediate scale, followed by a
strong rises towards still smaller scales. The minimum occurs on
scales that should formally still be well resolved, because these
scales are considerably larger than the maximum SPH smoothing length
$h_{\rm max}$ among all the particles (indicated as the vertical green
dotted line at $2\pi/h_{\rm max}$ in
Fig.~\ref{fig:sub:power256}). Nevertheless, already on this
comparatively large scale, the power starts to increase again. This is
a result of the high small-scale subsonic velocity noise present in
SPH that we already witnessed in Figure~\ref{fig:sub:velocity}. Even
when a kernel-smoothed velocity field is considered instead, and the
power spectrum is calculated for this smoothed quantity, there is a
considerable small-scale bump left, as shown by the dashed green line
which shows the power spectrum of the SPH-smoothed velocity field.  On
large scales, the behaviour of this field is the same as for the
nearest neighbour interpolated one, as expected.

\begin{figure}
\begin{center}
\setlength{\unitlength}{1cm}
\includegraphics{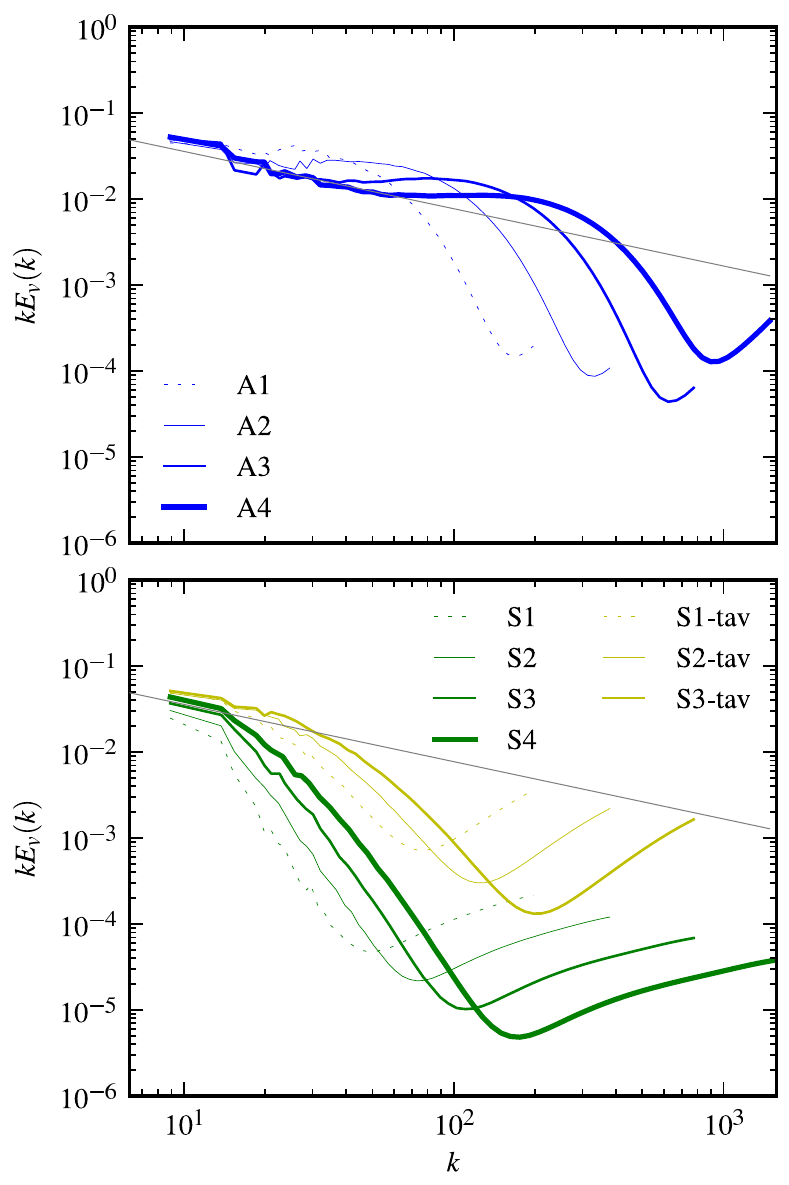}
\caption{Convergence study for the velocity power spectrum of
  ${\mathcal M}\sim \mach$ subsonic turbulence. The panel on top shows
  results for {\small AREPO}, from a resolution of $64^3$ to $512^3$
  cells. The panel on the bottom gives the corresponding results for
  SPH. However, even at a high resolution as high $512^3$ particles,
  no extended inertial range of turbulence can be identified in SPH.
  The thin grey lines show  the power-law expected for Kolmogorov's theory.
}
\label{fig:sub:powerRes}
\end{center}
\end{figure}

In Figure~\ref{fig:sub:powerRes}, we show a resolution study for the
subsonic velocity power spectra of our {\small AREPO} and SPH runs,
ranging from $64^3$ to $512^3$ particles/cells.  The SPH simulations
 seem to converge to each other only on the largest
scales. Even with a resolution as high as $512^3$ particles, there is
build up of an extended inertial range with the expected energy
cascade. We only see that with improving resolution there is a slight
shift towards smaller scales of the rapid decline of the power
spectrum. Also, the minimum of the power spectrum is reached at
progressively smaller scales, but the overall shape of the velocity
power spectrum does not improve significantly, and the small-scale
noise bump remains present.

For the simulations with {\small AREPO}, we observe that the
bottleneck effect moves to smaller scales with improving
resolution. This is expected, as this effect should be tied to the
numerical dissipation occurring on scales close to the resolution
limit. As the bottleneck moves towards smaller scales, a larger
inertial range with a self-similar power-law region is established on
large scales. We note that the rise of the power in the moving
mesh-code on very small scales, at around the Nyquist frequency, is
due to noise and aliasing effects at the spatial resolution limit that
is reached here, which is qualitatively a very different effect from
the small-scale velocity noise that sets in in SPH on much larger
scales.

\begin{figure}
\begin{center}
\setlength{\unitlength}{1cm}
\includegraphics{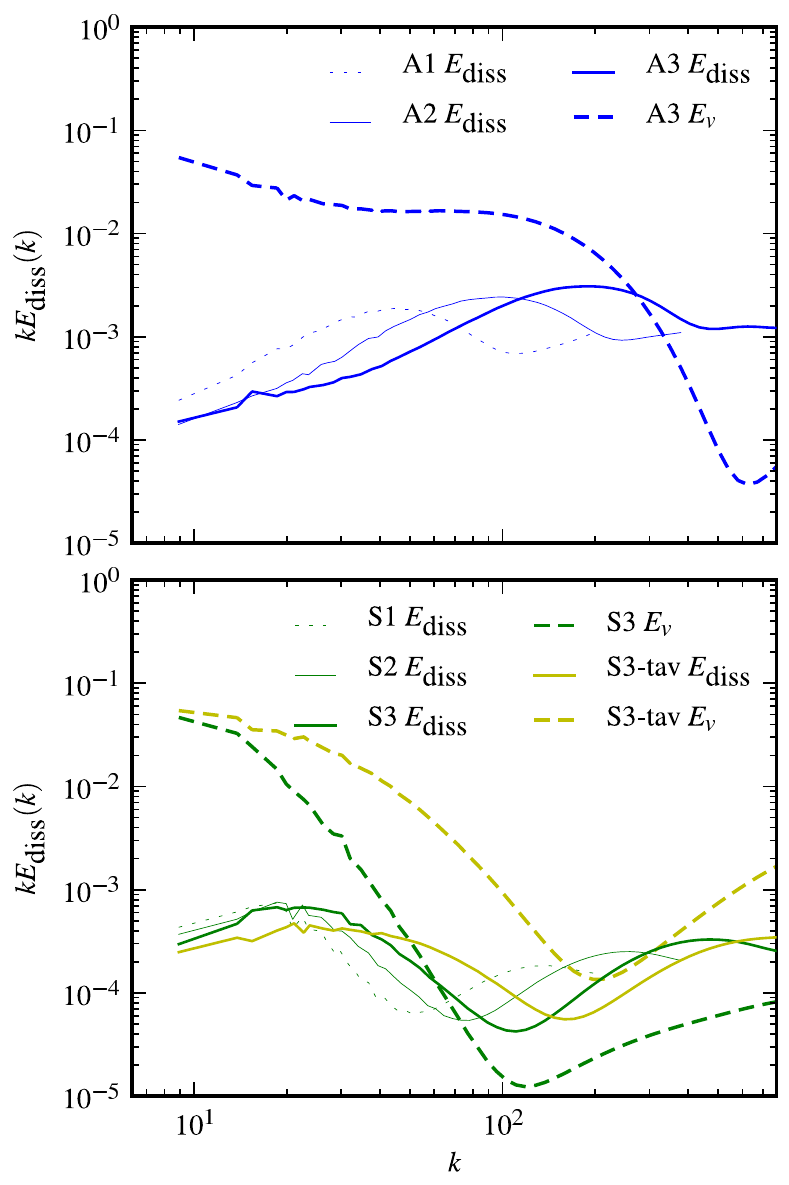}
\caption{Dissipation power spectra for {\small AREPO} and SPH runs at
  different resolutions, compared to the corresponding shape of the
  velocity power spectrum at $256^3$ resolution (dashed lines). For
  the mesh-code, the dissipation actually peaks on scales where the
  power spectrum starts to deviate from Kolmogorov's self-similar
  scaling. In contrast, SPH shows very strong dissipation already on
  larger scales, preventing the build-up of a turbulent cascade. In
  addition, the dissipation is also strong on small scales, close to
  the resolution limit, where the small-scale noise developing in SPH
  is constantly damped away.}
\label{fig:sub:powerDis}
\end{center}
\end{figure}

The computational cost of one of our moving-mesh turbulence
simulations with {\small AREPO} in the sub-sonic regime is nearly a
factor of $4$ higher than a corresponding SPH simulation with {\small
  GADGET-3} at the same number of resolution elements. Compared to a
corresponding fixed-mesh calculation, the moving-mesh simulation is
about $5$ times slower. This difference arises mainly due to the
costly Voronoi mesh construction, and in the case of moving versus
fixed mesh, additionally due to the roughly twice as many Riemann
problems that need to be solved for the unstructured Voronoi mesh
compared with a Cartesian grid.  However, these differences in run
time of order unity are implementation dependent and ultimately of
limited importance. What should really be considered is the
computational cost to reach a desired level of accuracy. For example,
as our moving mesh run with $128^3$ cells (A2) produces a model for
the Kolmogorov spectrum at least as good as the SPH run with $256^3$
particles (S3), one may state that {\small AREPO} is actually about
$4$ times as efficient as SPH when comparing these two runs. But we
caution that such statements about the relative efficiency of
different schemes are in general resolution and problem dependent. For
example, as we shall argue below (subsection \ref{subsecreynolds}), we
expect that the effiency gain of {\small AREPO} relative to SPH at
fixed accuracy actually grows with Reynolds number. Also, the relative
cost of moving-mesh vs. fixed-mesh depends on the Mach number of the
turbulence, because for highly supersonic flows considerably smaller
timesteps are needed for a fixed mesh compared with a moving mesh.
Finally, we remark that in astrophysical applications that include
self-gravity the run time difference between {\small AREPO} and
{\small GADGET} at the same number of resolution elements shrinks
considerably, simply because of the comparatively high cost of the
tree-based gravity calculation.

\subsection{Dissipation power spectra}

In Figure~\ref{fig:sub:powerDis}, we show power spectra for the energy
dissipation rate, measured as described in
Sections~\ref{sec:meth:diss} and \ref{sec:meth:power}. In the top
panel, we show results for the simulations A1 to A3 with resolutions
$64^3$ to $256^3$, averaged over the same period of time as in our
velocity power spectrum plots. For comparison, we also plot the
kinetic energy power spectrum as dashed lines, in order to allow a
comparison of the shapes of the different curves.  Interestingly, the
{\small AREPO} simulations show a peak in dissipation right at the
scales where the velocity power spectrum begins to rapidly fall.
While there is also some residual dissipation at very large scales
(which becomes smaller with better resolution), this is more than an
order of magnitude lower than the energy drained around the scales
where the dissipation measurement peaks. The result is hence
consistent with an interpretation where only negligible dissipation
occurs on large scales, with all the energy dissipated on some smaller
dissipation scale, which in our case is related to the numerical
resolution limit. Such a scenario is consistent with the theoretical
assumptions that enter Kolmogorov's theory of self-similar scaling.

In the bottom panel of Figure~\ref{fig:sub:powerDis}, we show the
corresponding SPH results.  Here a very different shape of the
dissipation power spectrum is found. There is a peak already on very
large scales, close to the driving scale. The amplitude of the
dissipation lies considerably higher on these scales range than in the
mesh-code, and shows only a weak dependence on numerical
resolution. This explains why there is not much energy left to be fed
into a turbulent cascade that could transport it conservatively
towards smaller scales.  Interestingly, there is however a second
extended maximum of the SPH dissipation power spectrum on very small
scales, coinciding with the location of the small-scale bump in the
velocity power spectrum. This is apparently related to viscous
dissipation of some of the small-scale velocity noise in SPH. 

\subsection{Dependence on SPH parameter settings}

\begin{figure}
\begin{center}
\setlength{\unitlength}{1cm}
\includegraphics{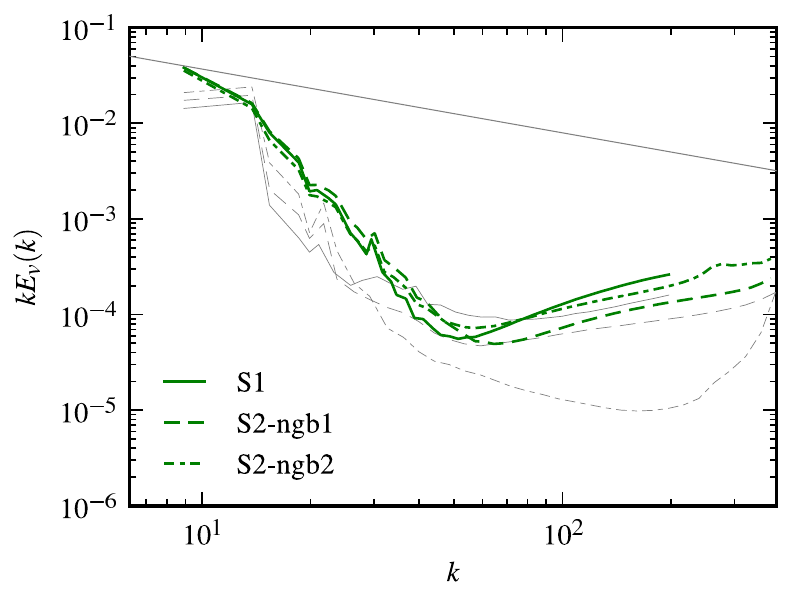}\\
\includegraphics{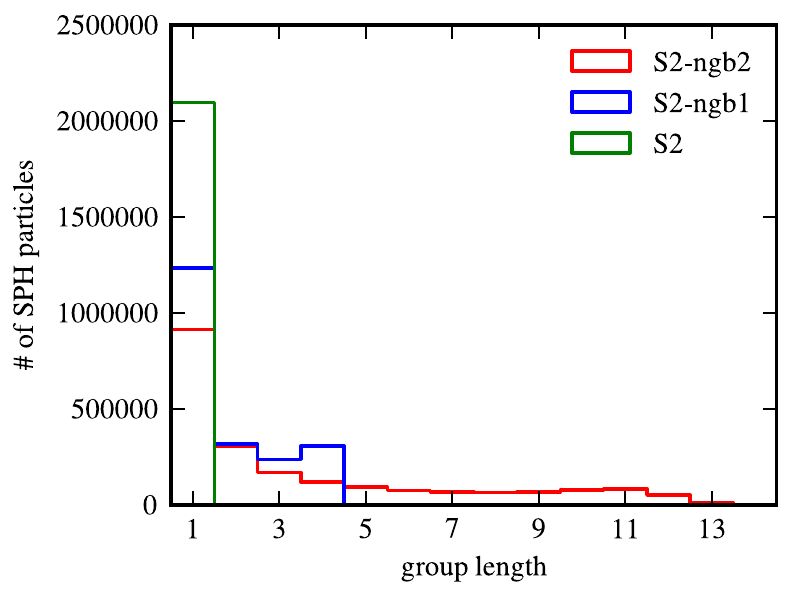}
\caption{ Effects of a larger number of SPH smoothing neighbours.  The
  panel on top gives results for the velocity power spectrum when the
  number of neighbours is increased from our default of 64 to 180, and
  finally to 512. Formally, the later run with $128^3$ particles has
  the same mass and spatial resolution as our S1 run with $64^3$
  particles. The lines shown in grey correspond to an early time, when
  the turbulence is not yet fully established. Here greater
  differences between the results can be seen, with the larger number
  of neighbours yielding clearly more power on large scales, and less
  power due to noise on small scales -- this is the expected effect of
  a better gradient accuracy due to a larger number of smoothing
  neighbours.  However, this advantage is quickly destroyed by the
  clumping instability. The bottom panel shows a histogram of SPH
  particle clump sizes determined with a friends-of-friends group
  finder, taking a linking length of 0.05 of the mean particle
  spacing. We see that the larger number of neighbours induces
  substantial clumping, reducing the number of independent sampling
  points and introducing large inhomogeneities in the kernel sums. }
\label{fig:sub:ngb}
\end{center}
\end{figure}

\begin{figure}
\begin{center}
\setlength{\unitlength}{1cm}
\includegraphics{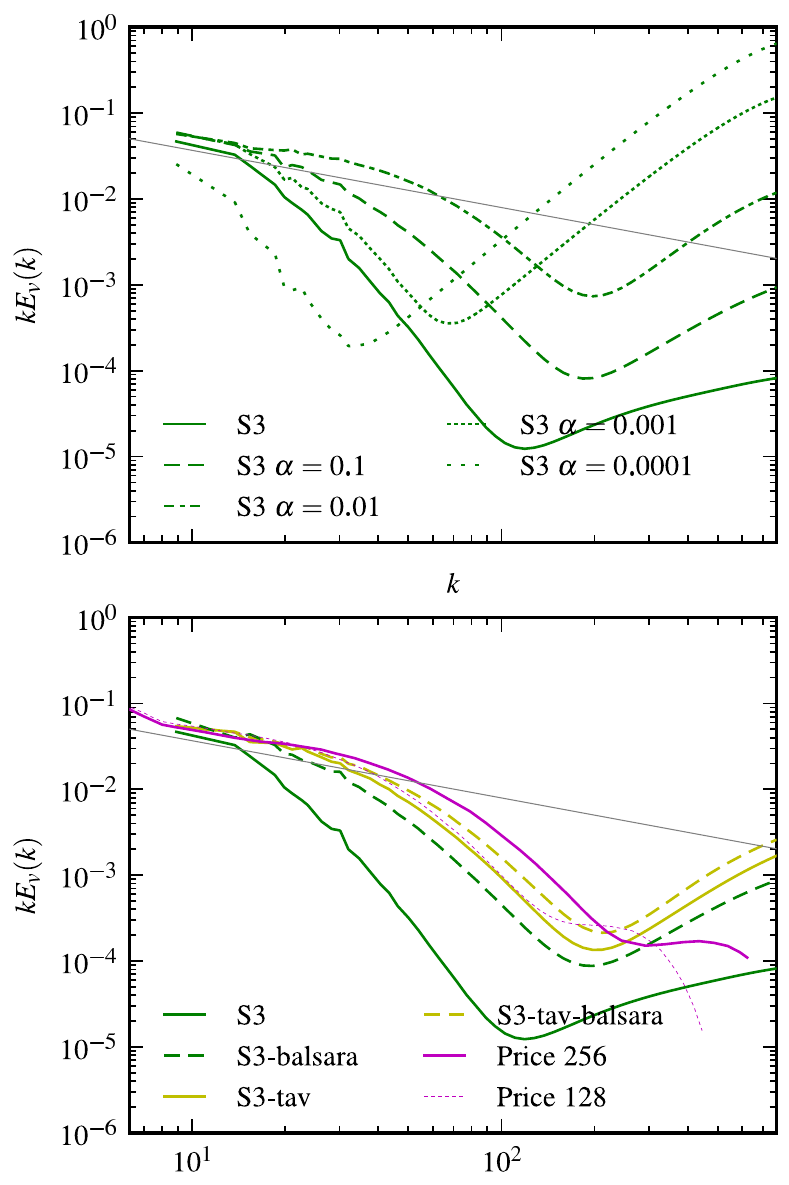}
\caption{ Dependence of SPH turbulence results on the viscosity
  parameterization.  The top panel illustrates the effect of
  systematically varying the SPH viscosity strength. For lower
  $\alpha$, the velocity power on large scales goes up, but the shape
  of the power spectrum does not improve.  Note however that this also
  increases the small scale velocity noise. However, if the viscosity
  strength is chosen too low, the power on large scales goes down
  again and the power spectrum is entirely noise-dominated in this
  case.  The bottom panel compares different artificial viscosity
  schemes. Enabling the Balsara viscosity suppression factor improves
  the power spectrum, yielding a result similar to our $\alpha=0.1$
  run.  A further improvement is achieved if a time dependent
  viscosity parameterization is applied. For comparison, we include
  the results of \citet{Price2012} for his $128^3$ and $256^3$ runs.
  The former is very close to our time-variable artificial viscosity
  run while the latter is comparable to our run with $\alpha=0.01$.
  The thin grey lines in both panels show the slope of the expected
  Kolmogorov power spectrum.  }
\label{fig:sub:powersph}
\end{center}
\end{figure}

Given the sobering results we have thus far obtained for subsonic
turbulence in SPH, it is an important question whether this outcome
can be significantly improved with different parameter choices for the
method. The primary numerical parameters that may strongly affect the
SPH results are the number of smoothing neighbours, and the artificial
viscosity parameterization. In fact, these are the only aspects that
can be changed easily without reverting to an entirely different
formulation of SPH, or a substantially different method for particle
hydrodynamics.

An increase in the number of smoothing neighbours should reduce the
noise in SPH kernel estimates. In fact, it has been argued that
convergence of SPH requires a simultaneous increase both of the number
of simulation particles and a (slower) increase of the number of
smoothing neighbours \citep{Rasio2000, Robinson2011}. Unfortunately,
in practice the clumping instability present for the normal SPH kernel
shape counteracts attempts to improve the SPH estimates through a
drastic increase of the number of smoothing neighbours \citep[but
  see][]{Read2010}. Regardless, we have examined whether an increase
of the number of neighbours to $N_{\mathrm{ngb}}=180$ or even
$N_{\mathrm{ngb}}=512$ improves our results. To this end we have
repeated our S2 simulation with these settings.

In the top panel of Figure~\ref{fig:sub:ngb}, we compare the velocity
power spectra of these two simulations with the S1 simulation.  Note
that at the resolution of $128^3$ employed for these tests, the S2 run
with 512 neighbours is expected to have effectively the same mass- and
spatial-resolution as the S1 simulation with our default choice of 64
smoothing neighbours.  Interestingly, the power spectra look very
similar on large scales, i.e.~there is no noticeable improvement due
to the higher number of smoothing neighbours at a fixed mass/spatial
resolution. Only the small-scale noise is reduced when the number of
neighbours is increased. However, if we look at early times after the
driving has started (grey lines in Fig.~\ref{fig:sub:ngb}), larger
differences are seen and the runs with a larger number of neighbours
do show more large-scale power, as expected for an improved accuracy.

The bottom panel of Fig.~\ref{fig:sub:ngb} highlights the reason why
this advantage soon vanishes. Here we determine the size spectrum of
particle clumps formed in the runs at the end of the simulated time by
applying a standard FOF algorithm with a small linking length of 0.05
times the mean particle separation. Whereas in the S2 simulation with
the default neighbour number essentially all particles stay isolated
and no groups are found, this is very different in the runs with
enlarged neighbour number. In the case of 512 smoothing neighbours,
less than half of the particles remain isolated, with a large number
of clumps containing multiple particles, up to $\sim 10$. This is the
well-known clumping instability that frustrates attempts to beat down
noise in the kernel sums by simply using a large number of smoothing
neighbours. Instead, the forming clumps reduce the effective
resolution and the accuracy of the kernel interpolants.

We now turn to the artificial viscosity parameterization, which is
another area where one may hope that simple changes could lead to
significant improvements in the results obtained for turbulence. In
particular, the problematic damping of the injected turbulent energy
already on large scales hints that a reduction of the viscosity may
help. A lower viscosity seems also warranted because in our subsonic
regime shocks are not really expected, suggesting that artificial
viscosity may perhaps not be needed at all, or only at a minimal
level. We have hence first repeated our default simulations with a
serious of reduced settings of the artificial viscosity parameter,
trying $\alpha=0.1$, $\alpha=0.01$, $\alpha=0.001$, and
$\alpha=0.0001$. The top panel of Figure~\ref{fig:sub:powersph} shows
the resulting velocity power spectra at S3 resolution. Compared to our
default S3 run, the large-scale power clearly increases when the
viscosity strength is reduced, but at the same time the small-scale
noise also drastically increased. In fact, we find that the energy
dissipated in this noise-dominated regime is essentially invariant
when the viscosity is varied (see Fig.~\ref{fig:sub:powerDis}). While
a larger artificial viscosity reduces the amplitude of the velocity
noise, it also implies stronger viscous forces, such that the average
work done against the viscous forces varies little.

 Eventually, however, the improvement of the large-scale results when
 lowering the viscosity ends. Instead the results deteriorate again
 when the viscosity is lowered to $\alpha=0.001$, or even
 $\alpha=0.0001$. In fact, for the latter case the large-scale result
 is even worse than for $\alpha=1.0$ whereas the small-scale noise is
 orders of magnitude larger. In this series of runs, it could be
 argued that $\alpha=0.01$ is ``best'' in the sense that the expected
 Kolmogorov slope of the velocity power spectrum is approximately
 reproduced over the widest range of scales. The large noise on small
 scales and the anemic `dip' in the power spectrum at $k\sim 200$,
 which falls short of the expected shape of the dissipation range (see
 below), raise significant concerns though that the statistical
 properties of these turbulent motions suffer from significant noise
 contaminations.

 In the bottom panel of Fig.~\ref{fig:sub:powersph}, we have instead
 enabled the so-called Balsara reduction factor for the viscosity in
 the presence of strong shear, and we consider a time-dependent
 viscosity parameterization. Again, both of these changes lead to an
 increase of the power on large scales and more noise on small scales
 compared to S3, as expected due to the reduced viscosity.  It turns
 out that the Balsara switch (S3-balsara) happens to lead to an
 extremely similar reduction of the effective viscosity as given by
 our run with $\alpha=0.1$ shown in the upper panel, but this is just
 by accident.  Compared to these two runs, the simulations with
 time-variable viscosity give slightly more power on all scales, and
 are hence still less viscous overall. Incidentally, the run without a
 shear viscosity limiter (S3-tav) agrees well with the result of
 \citet{Price2012} for his $128^3$ run with time-variable viscosity
 (shown also in the figure, for comparison), except on small scales
 due to Price's choice of measuring the power spectrum of the
 SPH-smoothed velocity field. However, the $256^3$ result of
 \citet{Price2012} with matching resolution lies still a bit higher,
 similar to our S3 result with $\alpha=0.01$. This is presumably
 caused my small differences in how the time-dependent viscosity is
 parameterized.

\begin{figure}
\begin{center}
\setlength{\unitlength}{1cm}
\includegraphics{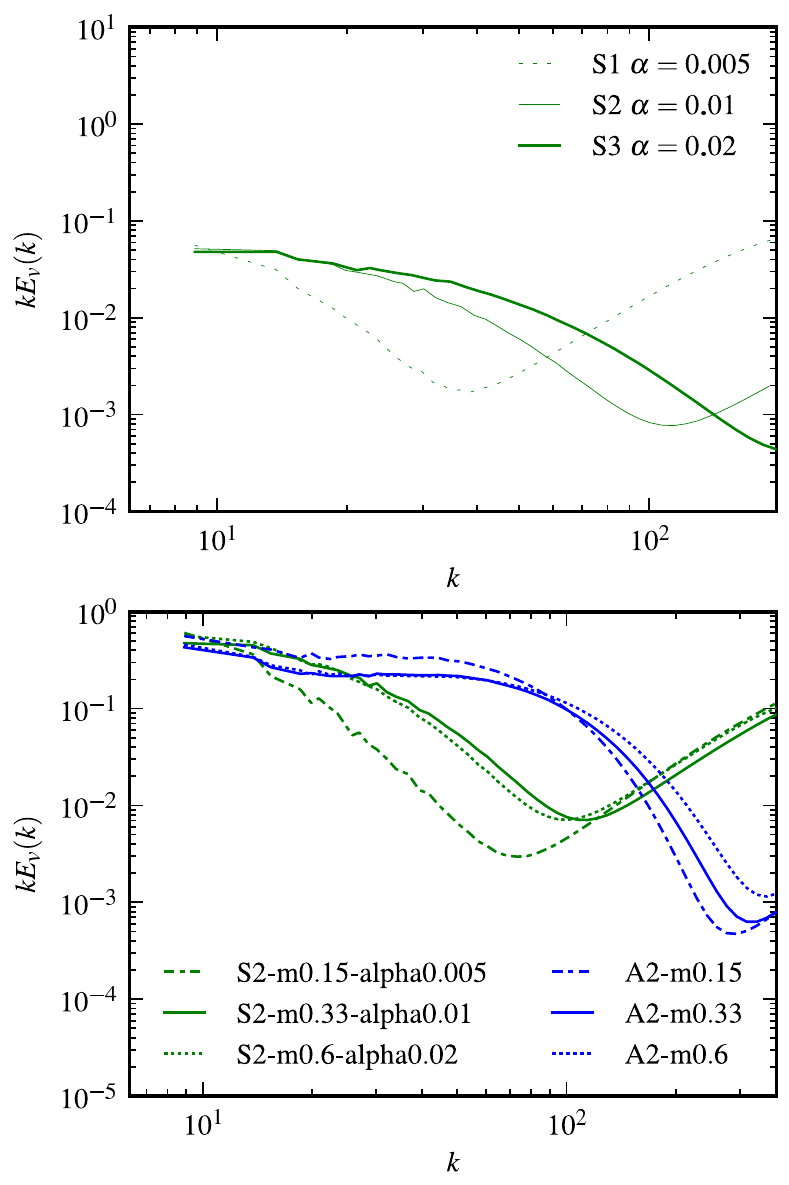}\\
\caption{ 
    Velocity power spectra in SPH for different numerical settings but
    equal Reynolds numbers. In the top panel, we compare results for
    different numerical resolutions, ranging from $64^3$ (S1) to
    $256^3$ (S3), and different artificial viscosity settings, such
    that the Reynolds number estimated according to equation (\ref{EqnRey})
    is the same and a very similar result would in principle be
    expected. In the bottom panel, we show instead results at a fixed
    resolution of $128^3$ (S2), but here the Mach number is varied and
    the viscosity parameter is adjusted such that the estimated
    Reynolds number stays the same. An approximate self-similarity of
    the spectral shape is at best obtained only for large Mach
    numbers. For comparison, we also show corresponding results for
    different Mach numbers for the moving-mesh code. Here the dynamic
    range of the inertial range is to a good approximation set by the
    grid resolution and is invariant with the Mach number.  }
\label{fig:sub:rey}
\end{center}
\end{figure}

\begin{figure}
\begin{center}
\resizebox{8.4cm}{!}{\includegraphics{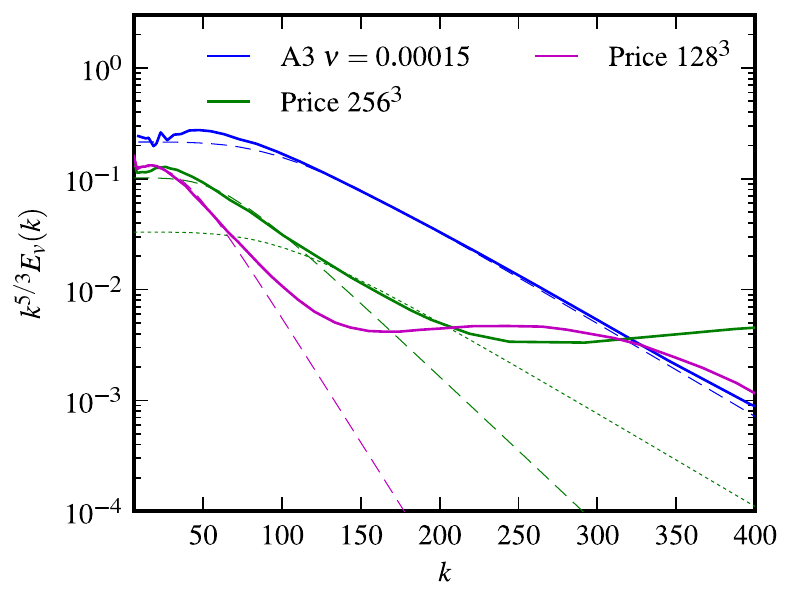}}
\caption{ Shape of the dissipation range in a moving-mesh simulation
  with physical viscosity, and in the SPH subsonic turbulence simulations
  of \citet{Price2012}. The dashed lines show fits to the velocity
  power spectrum based on equation (\ref{eqnfit1}), corresponding to Reynolds
numbers $2100$, $1000$ and $540$ for {\small AREPO} and the $256^3$ and
 $128^3$ SPH results, respectively. The green dotted line is the expected
shape for  ${\cal R}_e = 2100$ turbulence, shifted in amplitude to fit
part of the SPH result with $256^3$ particles. It is clear that the
dissipation range of the SPH results is not consistent in detail with
a Kolmogorov cascade. We note that \citet{Price2012} quotes a Reynolds
number of 6000 for his $256^3$ result.
}
\label{fig:sub:dissipation_range}
\end{center}
\end{figure}

\subsection{Reynolds numbers} \label{subsecreynolds}

In a response paper to our original submission the present work,
\citet[][]{Price2012} argued that our SPH results for subsonic
turbulence can be naturally explained simply by our artificial
viscosity settings. In fact, he suggests that SPH effectively yields a
solution of the Navier-Stokes equation with a kinematic viscosity
$\nu$ that can be estimated as
\begin{equation}
\nu \simeq \frac{1}{10}\alpha \,v_{\rm sig} h,
\end{equation}
where $\alpha$ is the artificial viscosity coefficient, $v_{\rm
  sig}\simeq c_s$ is the signal velocity, and $h$ is the SPH smoothing length. 
He further
argues that the Reynolds
number of the SPH turbulence simulations is then given by
\begin{equation}
{\cal R}_e \equiv \frac{L V}{\nu} = \frac{10\,L}{\alpha\,h} {\cal M}.
\label{EqnRey}
\end{equation}
Adopting as characteristic velocity scale the {\em rms}-velocity of the
turbulent velocity field (i.e. $V= {\cal M} c_s$), and for the length
scale $L$ the box size, \citet{Price2012} estimates a Reynolds numbers
of ${\cal R}_e \simeq 6000$ for his $256^3$ run. He also claims that
``turbulent flow with a Kolmogorov spectrum can be easily recovered''
in SPH when $\alpha$ is reduced away from shocks, and that our
argument that gradient errors are responsible for a flawed
Kolmogorov cascade would be ``incorrect''.
 
Actually, it is these assertions that are incorrect, as we show
now. If the Reynolds number of SPH was indeed simply given by equation
(\ref{EqnRey}) as suggested by \citet{Price2012}, we would expect {\em
  invariant} results (in the sense of Reynolds-number similarity) for
the turbulent power spectrum in different simulations when the
Reynolds number and the driving are kept constant. This is however not
the case, as is shown in Figure~\ref{fig:sub:rey}. Here we compare in
the top panel a series of different SPH simulations in which the
resolution is progressively increased by factors of 2, and the
viscosity parameter $\alpha$ is correspondingly changed such that the
(estimated) kinematic viscosity and Reynolds number stay fixed in all
cases. Nevertheless the numerical results for SPH do not line up.  In
the bottom panel of Fig.~\ref{fig:sub:rey}, we vary the Mach number of
the turbulence at fixed resolution, again adopting different viscosity
settings such that the Reynolds number according to
equation~(\ref{EqnRey}) should stay the same. Again, an accurate
universality of the shape of the SPH turbulence is not recovered,
although there is a hint that this may work better for large Mach
numbers.

Another important point to make is 
that not only the inertial range of
Kolmogorov turbulence for a Navier-Stokes flow is universal, the
dissipation range is universal as well \citep[e.g.][]{Pope2000}. In fact,
experiments demonstrate that the energy spectrum can be well described
by
\begin{equation}
E(k) = C \epsilon^{2/3} k^{-5/3} f_\eta(k\eta),  \label{eqnfit1}
\end{equation} 
where $\epsilon$ is the dissipation rate, and 
\begin{equation}
\eta \equiv  \left( \frac{\nu^3}{\epsilon}\right)^{1/4}
\end{equation}
is the Kolmogorov scale. The function $f_\eta(x)$ is universal and well fit by
\begin{equation}
f_\eta(x)= \exp\left(-\beta [(x^4 + c^4)^{1/4} - c]\right), \label{eqnfit2}
\end{equation}
with $c \sim 0.4$ and $\beta \sim 5.2$, and the value of the
Kolmogorov constant, $C\sim 0.5$, is universal as well. Note that this
means that for quasi-stationary turbulence, where the energy injection
rate is equal to the dissipation rate, not only the inertial range is
specified but also the shape of the spectrum in the dissipation range
is fully specified if the kinematic viscosity is known.

In Figure~\ref{fig:sub:dissipation_range}, we plot the velocity power spectra for
the $256^3$ and $128^3$ SPH runs of \citet{Price2012}, and we compare
them to a run with our Navier-Stokes solver in {\small AREPO}
\citep{Munoz2011}, with $\nu = 0.00015$ at a resolution of $256^3$
cells. The expected Reynolds number of this mesh-based simulation {\em
  with prescribed physical viscosity} (physical + numerical viscosity) is ${\cal R}_e\sim 2100$. By
plotting the spectrum using a linear scale for $k$, the dissipation
range is emphasized and becomes approximately a straight line. We
readily see that {\small AREPO} provides an excellent fit to the
expected shape (dashed line), as described by equations
(\ref{eqnfit1}) and (\ref{eqnfit2}), only a small deviation due to the
bottleneck bump is present.  However, neither the $128^3$ nor the
$256^3$ SPH results of \citet{Price2012} provide a reasonable fit to
the shape expected for Kolmogorov turbulence in the dissipation
range. This invalidates the claim that the subsonic turbulence results
of \citet{Price2012} (which are quite consistent with our own for
the similar viscosity settings, see Fig.~\ref{fig:sub:powersph}) are
consistent with Kolmogorov turbulence. It also shows that the Reynolds
numbers quoted by \citet{Price2012} are incorrect and too high by a
factor of $\sim 6$; in fact, his $256^3$ run has at most ${\cal R}_e
\sim 1000$, and the $128^3$ run about half that, but since the results
do not accurately correspond to the expected Navier-Stokes solutions
for these Reynolds numbers, these values have to be taken in any case
with a grain of salt.  We also want to remark that the dynamic range
expected for the inertial range of Kolmogorov turbulence is of order
$\eta/L \sim {\cal R}_e^{-3/4}$. A Reynolds number of 6000 should
hence allow up to $\eta/L \sim 680$ -- clearly infeasible with 256
points per dimension.

\begin{figure*}
\begin{center}
\setlength{\unitlength}{1cm}
\includegraphics{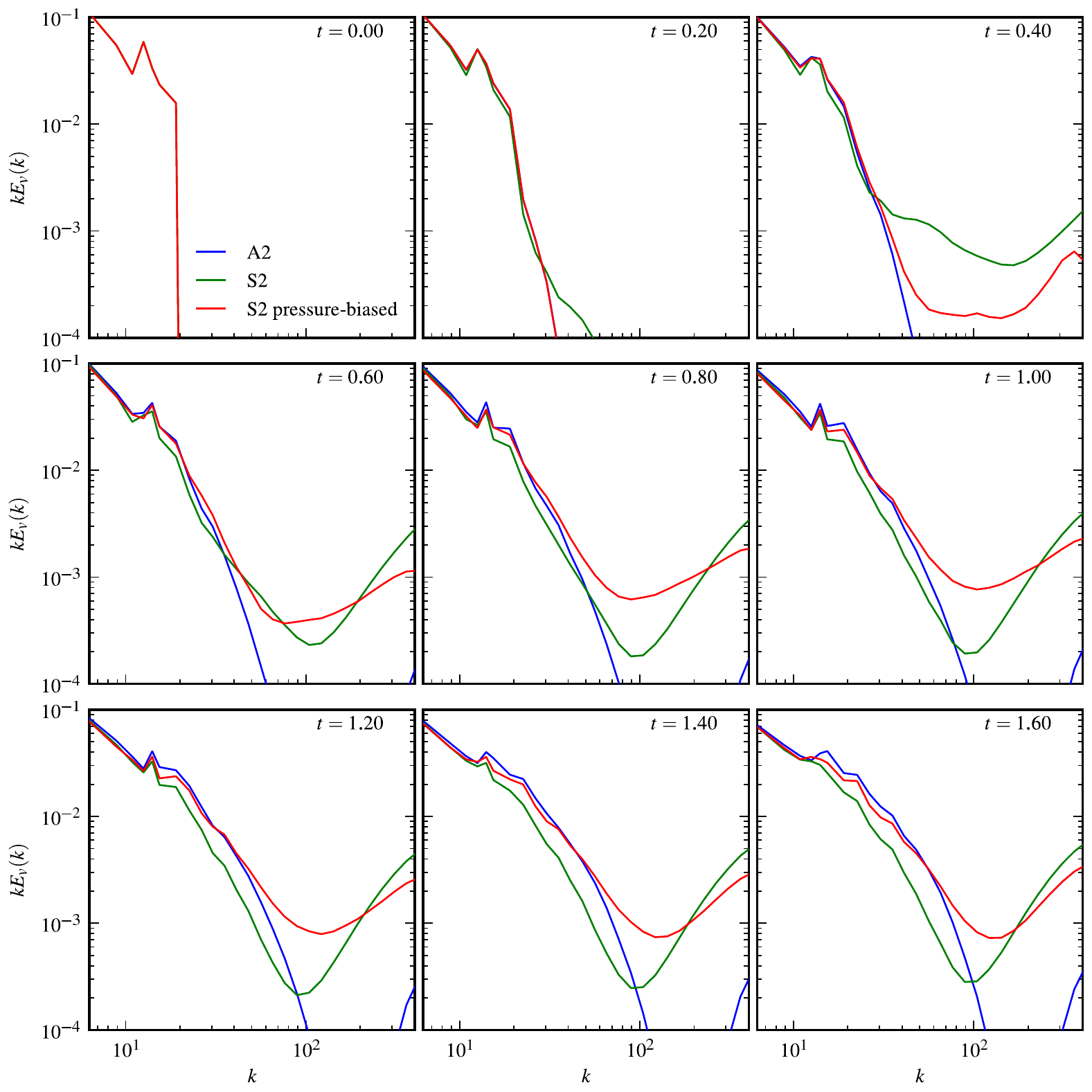}
\caption{Time evolution of the velocity power spectrum of a
  decaying large-scale solenoidal velocity field. Initially, only the
  $\sim 70$ largest modes are populated with random phases and an
  expected $E(k)\propto k^{-5/3}$ energy spectrum, normalized such
  that the {\em rms} velocity corresponds to ${\cal M}=0.3$.  We
  compare three different simulation techniques: {\small AREPO}
  (blue), ordinary SPH (green) with a low viscosity of $\alpha=0.01$,
  and ``pressure-biased'' SPH (red), where the only difference is that
  a constant pressure $P_0=\rho_0 c_s^2$ is subtracted in the equation
  of state. The resolution is $128^3$ in all cases. We see that
  ordinary SPH builds up small-scale noise considerably more quickly
  than the pressure-biased version of SPH. The latter actually tracks
  the {\small AREPO} result on large and intermediate scales much
  longer than ordinary SPH. As the viscosity in both SPH versions is
  the same, the faster dissipation of the large-scale motions in
  ordinary SPH is due to the larger noise on the smallest scales, as
  the energy dissipation rate $E_{\rm diss}(k)\propto k^2 E(k)$ is
  dominated by these scales.}
\label{fig:dis:decay}
\end{center}
\end{figure*}

Finally, we turn to the issue of gradient errors in SPH, which according to
\citet{Price2012} play no role in the turbulence results. To examine
this point, we carry out a simple experiment of a decaying large scale
solenoidal velocity field. We set up a simple random realization of a
number of solenoidal velocity modes on large scales (the largest 70
$k$-modes in the box), with a $\propto k^{-5/3}$ energy spectrum and
normalized such that the resulting {\em rms}-velocity corresponds
exactly to a ${\cal M}=0.3$ Mach number. We now compare the time
evolution of this field in three different simulations, without
applying any driving. We expect that the large-scale shearing motions
in the initial conditions will transfer some of their energy to
smaller scales with time, and that the damping of these motions by
numerical viscosity effects will eventually dissipate the kinetic
energy of the system.  One of the three codes we try is our
moving-mesh code {\small AREPO}. The second is standard SPH with a low
viscosity setting of $\alpha =0.01$. The third simulation uses the
very same SPH code, but with the only difference being that the
pressure in the equation of motion is replaced by $P \to P - P_0$,
where the constant $P_0=\rho_0 c_s^2 $ is taken to be the background
pressure. In principle, such a constant pressure offset should not
change anything, as pressure gradients are unaffected and hence no
effect on the dynamics is expected. However, SPH's equation of motion
has a `zeroth-order' error in the pressure gradient which is
proportional to the pressure itself \citep[e.g.][]{Quinlan2006,
  Read2010, Gaburov2011, Amicarelli2011}. For example,
\citet{Read2010} show that the actual acceleration acting on a SPH
particle corresponds to
\begin{equation}
\frac{{\rm d}\vec{v}_i}{{\rm d}t} = -\frac{P_i}{h\rho_i}\vec{E}_{0,i}
- \frac{(\vec{V}_i\nabla_i) P_i}{\rho_i} + {\cal O}(h) , \label{eqnE0}
\end{equation}
where $\vec{E}_{0,i}$ is a dimensionless error vector, and $\vec{V}_i$
is a dimensionless error matrix. Only for $\vec{E}_{0,i}=0$ and
$\vec{V}_i$ equal to the identity matrix the correct equation of
motion would be obtained. For an irregular particle distribution, one
obtains however $\vec{E}_{0,i}\sim {\cal O}(h)$, meaning that the
zeroth-order error is not easily reduced with better resolution.
However, by subtracting $P_0$, the average pressure of a particle is
made close to zero, and with this trick the magnitude of the
zeroth-order error should be greatly reduced.

Indeed, if we look at the time evolution of the power spectrum of our
decaying velocity field (Figure~\ref{fig:dis:decay}), we see a
substantial difference between the two flavours of SPH. The standard
version of SPH builds up small-scale velocity noise much more quickly,
and the power on large scales decays more rapidly as well. In
contrast, the version of SPH with a pressure-bias in the equation of
state manages to track the {\small AREPO} result on large scales for a
much longer time. This hence proves that the small-scale noise created
by gradient errors drains kinetic energy from large scales,
effectively short-circuiting the Kolmogorov cascade.  Notice that the
two flavours of SPH examined here have identical viscosity settings,
i.e.~their only difference lies in the gradient errors in the SPH
equation of motion. Unfortunately, in general applications of SPH in
astrophysics the subtraction of a constant background pressure is not
readily possible, so the problem of gradient errors is present
irrespective of the artificial viscosity settings.

We note that the problematic influence of noise and gradient errors in
SPH has also become apparent in recent tests of the Gresho vortex
problem \citep{Springel2010b, Read2011}. The stable vortex flow in
this set-up \citep{Gresho1990} tends to decay relatively quickly in
standard SPH, but \citet{Read2011} recently showed that the error and
the convergence rate can be greatly improved if a higher-order
gradient estimate, based on a much larger number of SPH smoothing
neighbours and a different kernel shape, is used.

\begin{figure}
\begin{center}
\setlength{\unitlength}{1cm}
\includegraphics{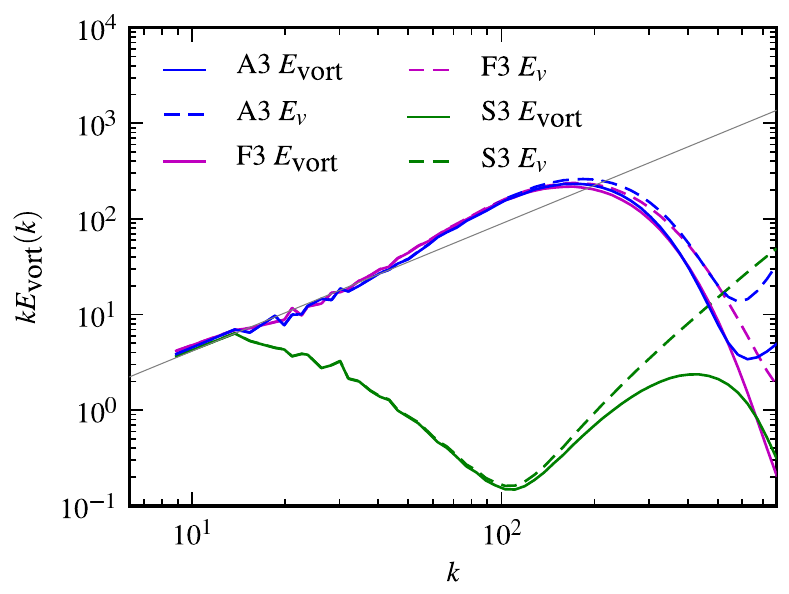}
\caption{Vorticity power spectrum of SPH and {\small AREPO}, compared
  at a resolution of $256^3$.  The thin grey line gives the slope
  expected for Kolmogorov's theory of incompressible turbulence, in
  which for fully developed turbulence, the power spectrum of the
  vorticity is proportional to $k^2$ times the velocity power
  spectrum. The {\small AREPO} result follows this expectation very
  well on large scales, over the same range where also Kolmogorov's
  velocity power spectrum is reproduced. In contrast, SPH shows a
  rapid fall of vorticity towards small scales; only on scales of
  order the SPH smoothing length a substantial vorticity bump is
  seen, but this is presumably largely due to the velocity noise on
  these scales.}
\label{fig:sub:vortspect}
\end{center}
\end{figure}

\begin{figure}
\begin{center}
\setlength{\unitlength}{1cm}
\includegraphics{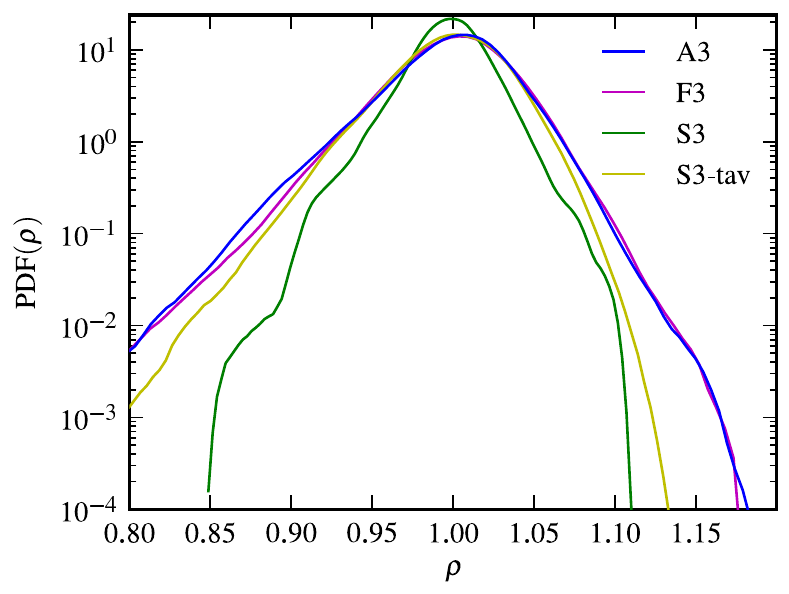}
\caption{Volume-weighted density PDFs for subsonic turbulence. Shown
  are SPH, {\small AREPO}, and fixed-mesh simulations. The PDFs
  are averaged over 5 snapshots taken at times $t=12.8$, 16.0, 19.2, 22.4
  and 25.6. }
\label{fig:sub:densitypdf}
\end{center}
\end{figure}

It is also interesting to consider the computational cost required to
reach a certain effective Reynolds number in SPH and in a mesh
scheme. If we ignore the issue of gradient errors in SPH for the
moment and assume that the Reynolds number is indeed given by equation
(\ref{EqnRey}), then the computational cost to reach a certain
Reynolds number scales as $t_{\rm CPU} \propto {\cal R}_e^4$. This is
because the number of particles scales as $h^{-3}$, and the number of
required timesteps as $h^{-1}$. In contrast, because in a mesh code
such as {\small AREPO} the Kolmogorov dissipation scale is essentially
given by the cell size, $\eta \sim h$, we expect ${\cal R}_e \propto
h^{-4/3}$, implying that the required CPU-time to reach a certain
Reynolds number scales only as $t_{\rm CPU} \propto {\cal R}_e^3$. In
the limit of large Reynolds numbers this is a significant competitive
advantage.

\subsection{Vorticity power spectrum}

Another interesting probe for turbulence is provided by the vorticity
$\vec{w} = \nabla \times \vec{v}$ of the velocity field. This is
because vorticity is in principle a conserved fluid quantity in an
ideal gas, where the vorticity field is locked into the gas similar to
a flux-frozen magnetic field in ideal magnetohydrodynamics. New
vorticity is introduced on large scales through our solenoidal driving
field, and it is erased on small scales via dissipation, but vorticity
production through the baroclinic term should essentially be absent in
our flow due to our quasi-isothermal conditions. Analyzing the
statistical properties of the vorticity field can hence provide
complementary information about turbulence and the numerical
properties of the employed simulation technique.

In Figure~\ref{fig:sub:vortspect}, we show our measurements for the
vorticity power spectrum for the same simulations that we used in
Fig.~\ref{fig:sub:power256} for an analysis of the velocity power
spectrum. Both our moving-mesh and fixed-mesh simulations show
approximately a power-law rise of the vorticity with scale, until a
rapid drop sets in at around the numerical dissipation scale. For
fully developed isotropic turbulence we expect a power-law spectrum of
the vorticity given by $E_{w} \simeq k^2 E_v$
\citep[e.g.][]{Zhu2010}. In particular, Kolmogorov's theory then
suggests a slope that is 2 units shallower than that of the velocity
spectrum, and hence rises towards smaller scales as $E_w \propto
k^{1/3}$. This expectation is borne out well by our measurements,
providing an important consistency check. Also, the proportionality
between $E_{w}$ and $k^2 E_v$ is resolved quite well, as shown by the
dashed lines.

However, as already expected based on the enstrophy maps shown in
Fig.~\ref{fig:sub:velocity}, the SPH results for the vorticity power
spectrum are very different. In essence, only some large eddies
resulting from the driving are present, apart from some vorticity
power on small scales, which is likely in large part noise, and in any
case is smaller than what is found in the mesh code. This hence
corroborates our previous results, confirming that the mesh-based
simulations do resolve subsonic turbulence with the expected physical
properties whereas this is more problematic in the case of SPH.

\begin{figure*}
\begin{center}
\setlength{\unitlength}{1cm}
\includegraphics{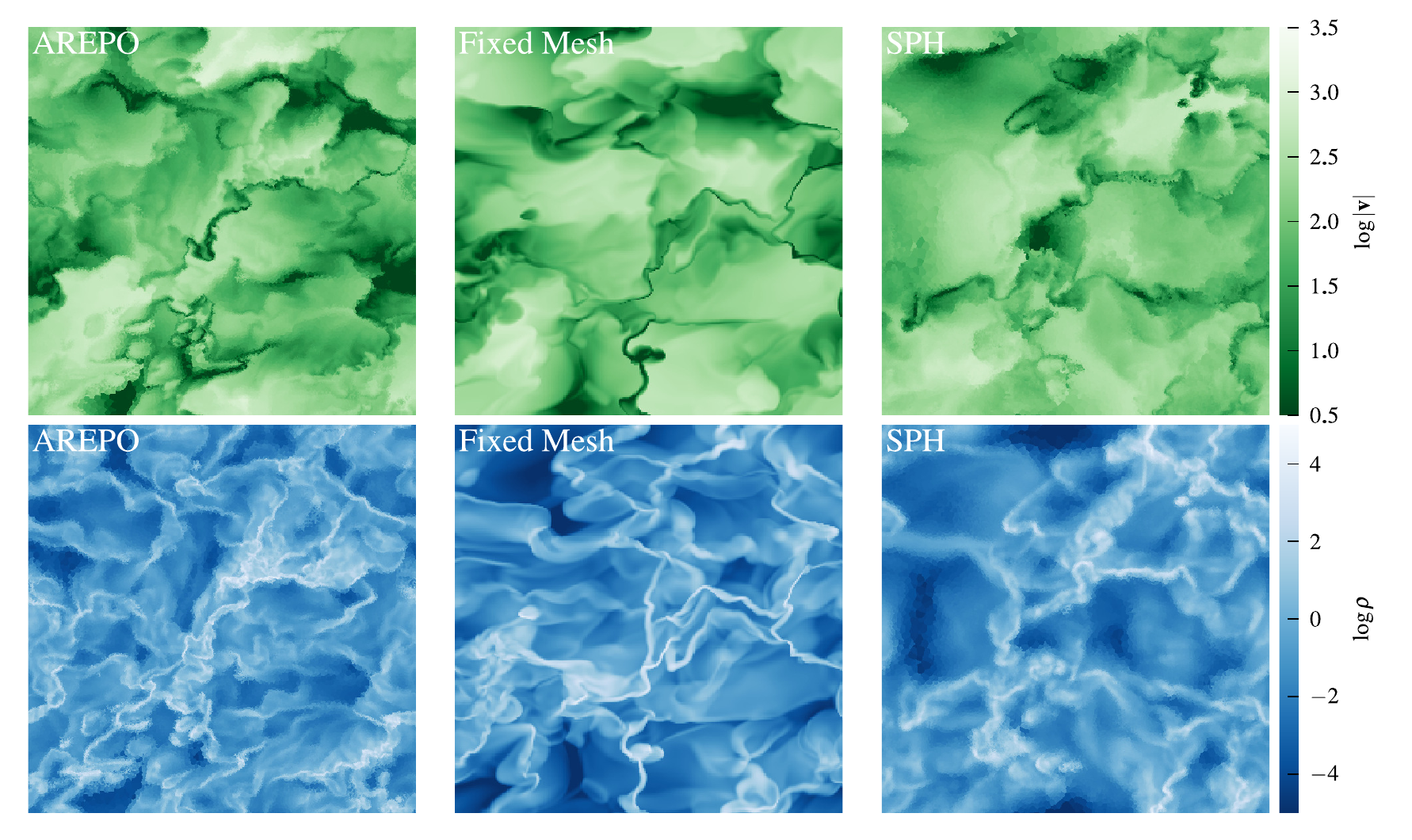}
\caption{Visualizations of the state of the gas in slices through our $\mathcal{M} \sim \mhigh$
  supersonic runs at $256^3$ resolution. The top row shows the
  logarithm of the velocity field and the bottom row the logarithm of
  the density field. From left to right, simulations with a moving
  mesh in {\small AREPO} (A3-m10), a fixed mesh (F3-m10), and SPH (S3-m10) are
  shown. \label{fig:supersonicslice}}
\end{center}
\end{figure*}

\begin{figure*}
\begin{center}
\setlength{\unitlength}{1cm}
\includegraphics{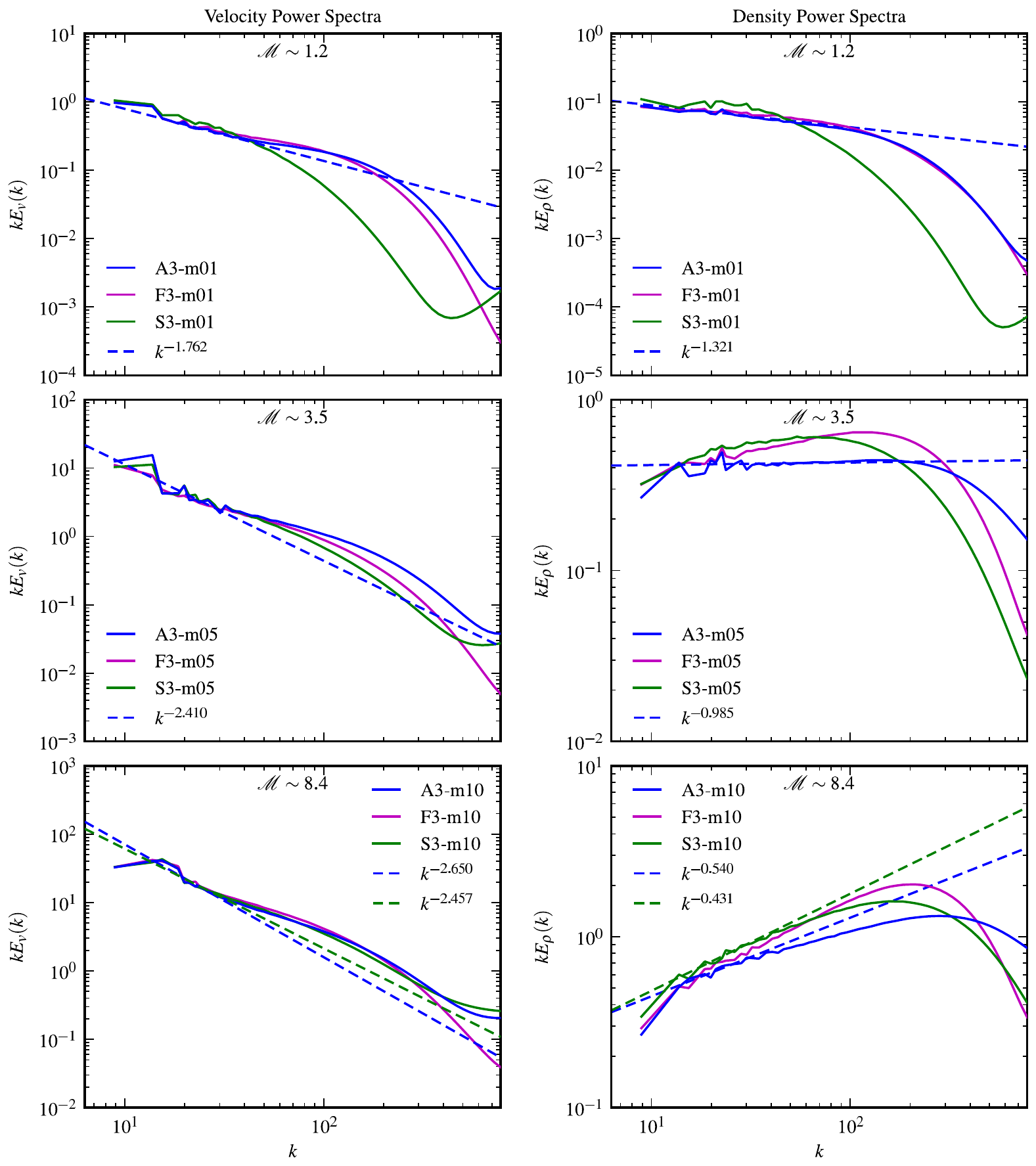}
\caption{Turbulence power spectra of SPH, {\small AREPO}, and a
  fixed-mesh code for approximately sonic (top panel), mildly
  supersonic (middle panel) and for highly supersonic driven
  isothermal turbulence (bottom panel). The left panels show the
  velocity power spectra, the right panels the density power spectra.
  The dashed lines show fitted power-laws for {\small AREPO} (blue) and SPH (green).
}
\label{fig:super:power}
\end{center}
\end{figure*}

\subsection{Density probability distribution function}

The final quantity we examine in our simulations of subsonic
turbulence is the density distribution function.  In
Figure~\ref{fig:sub:densitypdf}, we compare the volume-weighted
density PDFs for our A3, F3, S3, and S3-tav simulations.  These
distribution functions have been averaged over 5 simulation snapshots
evenly spaced in time between $t=12.8$ and $t=25.6$ since the start of
the simulations.

We find that the mesh codes agree well in their density PDF. However,
the {\small AREPO} code in moving-mesh mode shows slightly more
high-density regions than when a fixed-mesh is used. This is
consistent with our expectation that the moving mesh yields a slightly
lower numerical viscosity and a better adaptivity to high density
regions. Due to the subsonic conditions in these simulations, the
differences are expected to be quite small though. The density PDF of
the SPH simulation shows somewhat larger differences. The
comparatively viscous $\alpha=1.0$ result is slightly thinner and
hence more strongly peaked towards the mean value than for the
mesh-based results. This is despite the fact that noise in the SPH
density estimates can be expected to broaden the intrinsic
distribution, but apparently this effect is negligible compared to the
physical width of the distribution one expects for $\mathcal{M}\sim
\mach$ turbulence. Interestingly, our SPH run with time-variable
artificial viscosity produces a density PDF very close to the
mesh-based results on the low-density side, whereas it falls short a
bit in the tail of the high-density side.

\begin{figure*}
\begin{center}
\setlength{\unitlength}{1cm}
\includegraphics{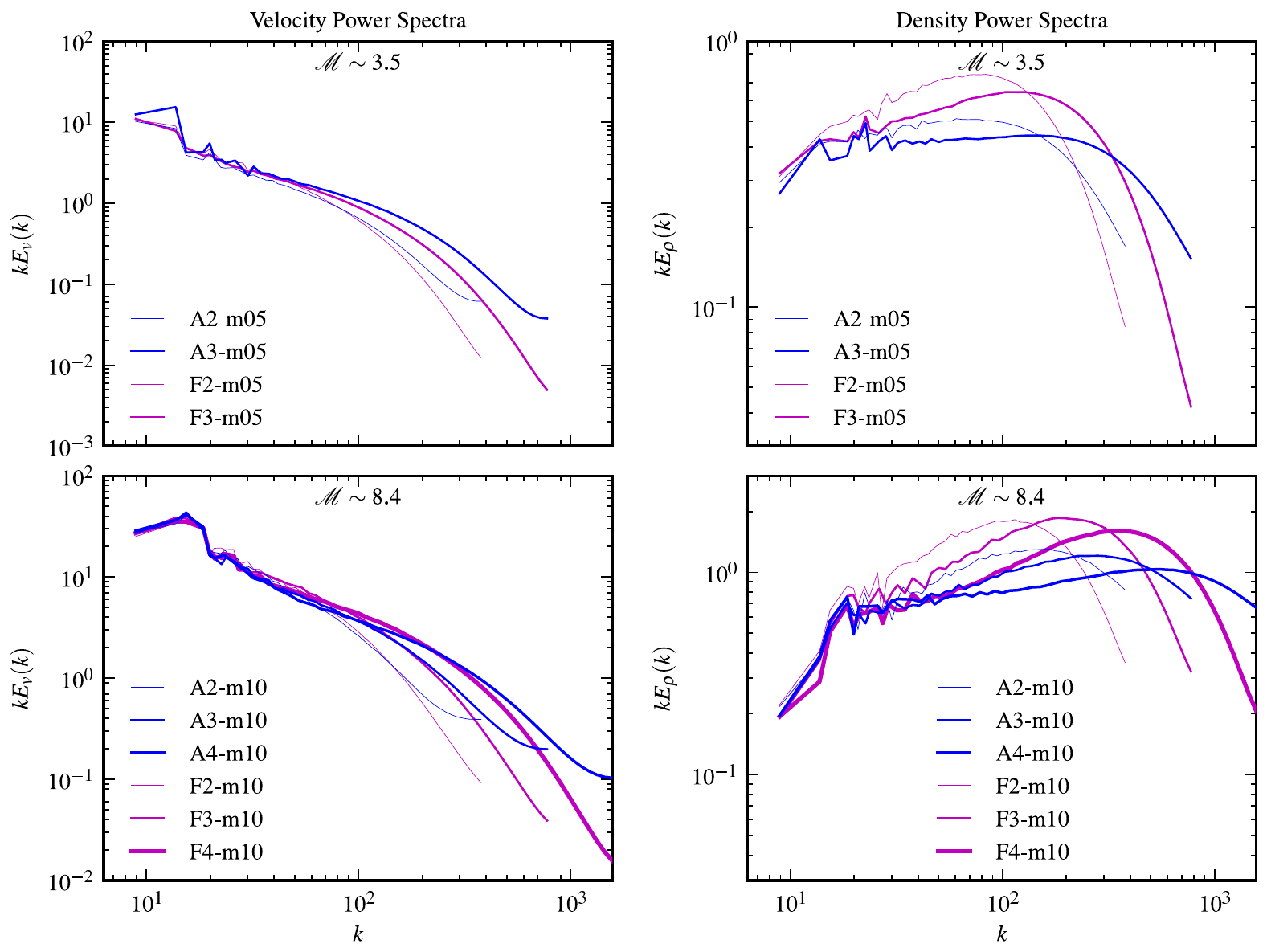}
\caption{Resolution study of the supersonic velocity and density power
  spectra for our fixed-mesh and moving-mesh simulations. The top
  panels show the runs at Mach number $\mathcal{M} \sim \mmed$, and
  the bottom ones at $\mathcal{M} \sim \mhigh$. At large and
  intermediate scales, the moving-mesh runs correspond roughly to the
  fixed-mesh runs at $2^3$ times higher resolution.}
\label{fig:super:densityres}
\end{center}
\end{figure*}

\section{Transsonic and Supersonic turbulence}  \label{sec:super}
 
Superficially, the results obtained thus far seem to be in conflict
with previous reports that SPH can adequately represent highly
supersonic isothermal turbulence. However, it is important to
appreciate that the physical nature of supersonic isothermal
turbulence is really quite different from the subsonic regime studied
thus far, both in terms of the role of ram pressure versus thermal
pressure, and in terms of the relevant dissipation mechanisms. In
supersonic turbulence, we expect a network of shocks, which in the
limiting case of fully kinetic turbulence is described by Burgers
turbulence and not by the Kolmogorov theory. 

We here briefly examine how well our new moving-mesh code {\small
  AREPO} describes turbulence in the transsonic and highly supersonic
regimes, and whether the SPH results improve in this regime. The names
and key parameters of the primary simulations we have carried out for
these tests are listed in Table~\ref{tab:supersim}.  A first
impression is obtained by the maps in
Figure~\ref{fig:supersonicslice}, where we show slices through the
density and kinetic energy density fields in our $\mathcal{M}=\mhigh$
simulations using a moving mesh in {\small AREPO}, a fixed mesh, or
SPH. Right away we notice a much greater similarity of the maps than
found in the subsonic regime, with SPH apparently being able to
resolve the turbulence in a way that is at least qualitatively similar
to the mesh-based results.

We examine this more quantitatively in Figure~\ref{fig:super:power},
where we show the velocity and density power spectra for increasing Mach
numbers, ranging from the transsonic to the supersonic regime.  We
include results for all three types of numerical simulations that we
have carried out, comparing them always with an identical forcing
field as a function of time.

The simulations shown in Fig.~\ref{fig:super:power} were performed for
Mach numbers $\mathcal{M} \sim \mtrans$ (top row), $\mathcal{M} \sim
\mmed$ (middle row) and $\mathcal{M} \sim \mhigh$ (bottom row). We
clearly see that as the Mach number increases, the SPH method does
progressively better for the velocity power spectrum and in fact
appears to eventually converge to the result obtained with the two
mesh-based techniques. While for ${\cal M} \sim \mtrans$, there is still
a very significant deficit of power in SPH except for the largest
scales, this effect becomes significantly weaker for $\mathcal{M}
\sim \mmed$ and almost vanishes for $\mathcal{M} \sim \mhigh$. When
one compares the velocity power spectra of the moving-mesh and the
fixed-mesh calculations, one generally finds very good agreement on
large scales but a noticeable difference in the small-scale
behaviour. The dissipation scale of the moving-mesh code lies at
slightly smaller scale, which can be interpreted as a signature of a
higher effective resolution for the same number of fluid cells. This
advantage most likely arises from the reduced advection errors in the
moving-mesh approach.

If instead the density power spectra are compared (right column in
Fig.~\ref{fig:super:power}), we find a qualitatively similar
behaviour. For higher Mach number, the SPH result approaches that of
the mesh-based simulations. There are however somewhat larger residual
differences between all three techniques at the highest Mach number
compared with the situation found for the velocity power spectrum. In
particular, the density power spectrum for the moving-mesh code has a
shallower slope and extends to higher $k$ than both for the fixed-mesh
and the SPH codes. We interpret this as a result of the better
adaptive resolution of the moving-mesh technique. Direct fits to the
power-law region of the density power spectra at $\mathcal{M} \sim
\mhigh$ return slopes of $-0.54$ and $-0.43$ for these $256^3$
moving-mesh and SPH runs, respectively, clearly indicating a
significant difference. However, we note that the shape of the density
power spectrum is relatively sensitive to resolution, as we show next,
so these slopes are not numerically converged.

In Figure~\ref{fig:super:densityres}, we show a resolution study for
our moving-mesh and fixed-mesh simulations for the cases of
$\mathcal{M} \sim \mmed$ and $\mathcal{M} \sim \mhigh$ turbulence,
considering results both for the velocity and density power
spectra. It is seen that the velocity power spectra agree nicely
between the moving-mesh and the fixed-mesh code at large scales, but
that the effective resolution of the moving-mesh code is higher at a
given number of resolution elements. When the resolution is improved,
the power-law region corresponding to the inertial range is extended
towards smaller scales, without a significant change in slope. A small
bottleneck effect still seems to be present, but at a much smaller
level than in the subsonic regime.

In contrast, convergence in the density power spectrum is more
challenging, as seen in the right column panels of
Fig.~\ref{fig:super:densityres}. Here low resolution can easily lead
to an overestimate of the slope due to a fairly prominent bottleneck
effect. Interestingly, in the $\mathcal{M}=\mhigh$ case we see that
the slope of the fixed-mesh F4-m10 simulation is accurately reproduced
already by the A3-m10 simulation, at an almost one order of magnitude
smaller number of resolution elements, and similarly for the F3-m10
and A2-m10 pair of simulations. This can be attributed to the adaptive
nature and the lower advection errors of the moving-mesh approach
compared with the fixed-grid Eulerian method.

\begin{figure}
\begin{center}
\setlength{\unitlength}{1cm}
\includegraphics{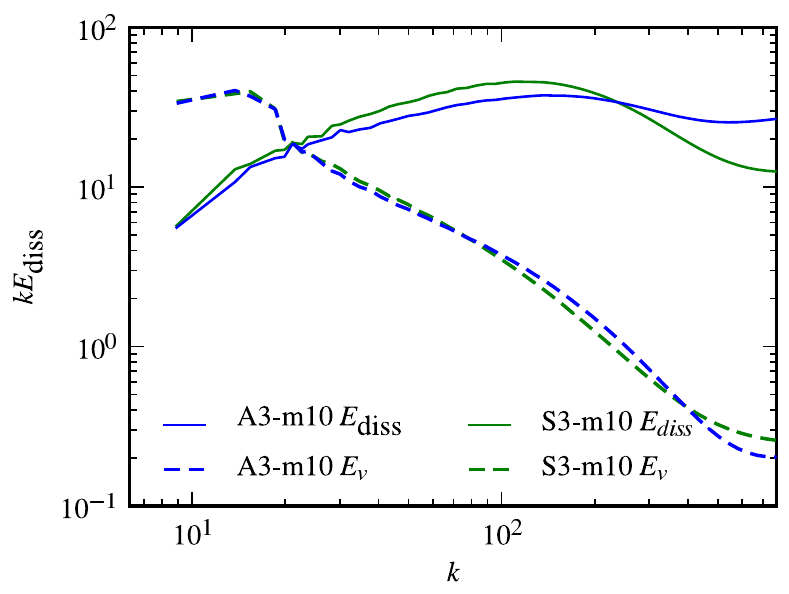}
\caption{Dissipation power spectra for the supersonic runs A3-m10,
  F3-m10 and S3-m10 at $\mathcal{M} \sim \mhigh$. As in Fig.~\ref{fig:sub:powerDis} the velocity
  power spectra are plotted as dashed lines.
}\label{fig:super:dis}
\end{center}
\end{figure}

\begin{figure}
\begin{center}
\setlength{\unitlength}{1cm}
\includegraphics{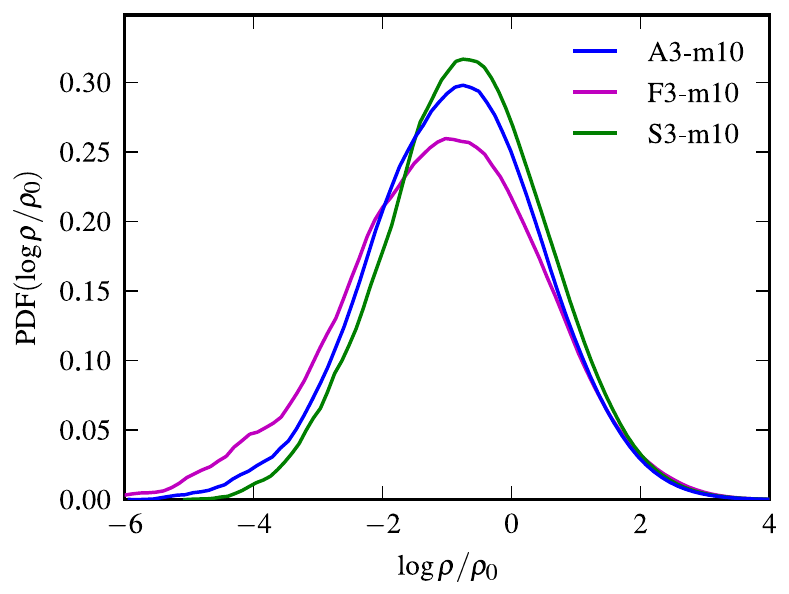}
\caption{The volume-weighted logarithmic density PDFs for our highly
  supersonic runs at $\mathcal{M} \sim \mhigh$, as labeled. The PDF is averaged over two snapshots
  at times $t=0.2$ and $t=0.3$. }
\label{fig:super:pdfsuper}
\end{center}
\end{figure}

It is also interesting to examine the energy dissipation power spectra
of the different simulation techniques in the highly supersonic
regime. This is done for our highest Mach number run in
Figure~\ref{fig:super:dis}. All the simulations show a relatively
broad distribution of dissipation as a function of scale, which is
quite different in character compared to the narrower distribution
encountered in the subsonic case. This can be interpreted as a result
of the different physical nature of the dissipation in this supersonic
regime, which occurs primarily through a complex network of shock
surfaces, and is hence not restricted to a small range of length
scales.  It is also interesting to note that SPH and the moving-mesh
code show quantitatively a quite similar result, whereas the
fixed-mesh method gives higher dissipation on nearly all scales. We
argue that this is due to the significant bulk motion present in the
system, inducing enhanced dissipation through advection errors in the
fixed-mesh code. This is simply not present in this form in the two
Lagrangian methods, which are both Galilean-invariant schemes.

Finally, Figure~\ref{fig:super:pdfsuper} takes a look  at the
volume-weighted density probability distribution function (PDF) in the
high Mach number case. We compare the PDFs of moving-mesh, fixed-mesh
and SPH simulations at the $256^3$ resolution.  The shape of all three
results is described reasonably well by a log-normal
distribution. However, the fixed-mesh simulation shows a higher
probability at the low density end and has the largest width of the
distribution for this reason.  The SPH simulation tends to give higher
probability at the high density end, which is a very similar behaviour
as found in \cite{Price2010}.  The moving-mesh run has an overall very
similar distribution as the SPH run, except for being slightly wider.
To the extent that a better representation of the high-density tail is
advantageous in science applications of supersonic turbulence (which
can be argued is particularly true in studies of star formation), the
moving-mesh technique hence works at least as well as SPH, and arguably
better than a fixed-mesh technique.

\section{Discussion and Conclusions} \label{sec:disc}

Perhaps the most important question prompted by our results is why SPH
behaves so badly in the subsonic regime.  \citet{Price2012} argued
that the culprit lies simply in the artificial viscosity
parameterization, and that schemes that dynamically reduce the
viscosity away from shocks
\citep[e.g.][]{Morris1997,Dolag2005,Cullen2010} do much better and
have no problem to reproduce a Kolmogorov cascade also in the subsonic
regime. While we agree that excessive artificial viscosity can
compromise the results of SPH, particularly in the subsonic regime
where this will show up more readily, this is by no means the complete
story. Instead, we have demonstrated in this study that gradient
errors inherent in standard formulations of SPH (`classic SPH') do
play a major role as well. They seed small-scale velocity noise in
shear flows on all scales, and this noise is dissipated away by the
viscosity of the scheme. Since SPH is an energy-conserving scheme,
this effectively short-circuits part of the energy transfer cascade in
Kolmogorov's theory of turbulence.

The concern that the large subsonic noise in SPH may cause substantial
accuracy problems in the treatment of fluid instabilities has recently
been emphasized by a number of authors \citep{Springel2010b, Read2010,
  Abel2011}. Also, numerous studies have pointed out that the standard
approach followed in SPH for constructing derivatives of smoothed
fluid quantities involves rather large error terms, especially for the
comparatively irregular particle distributions in multi-dimensional
simulations. One problem lies in a lack of consistency of the
ordinary density estimates (which do not conserve volume, i.e.~the sum
of $m_i/\rho_i$ is not guaranteed to add up to the total volume), and
another in a low order of the gradient estimate itself
\citep[e.g.][]{Quinlan2006, Read2010, Gaburov2011, Amicarelli2011}.
In practice, this means that there can be pressure forces on particles
even though all individual pressure values of the particles are equal,
a point emphasized in a recent study by \citet{Abel2011}. But if this
is the case, spurious jittering motions of particles can be readily
triggered even for a vanishingly small large-scale pressure gradient.

\begin{figure}
\begin{center}
\setlength{\unitlength}{1cm}
\includegraphics{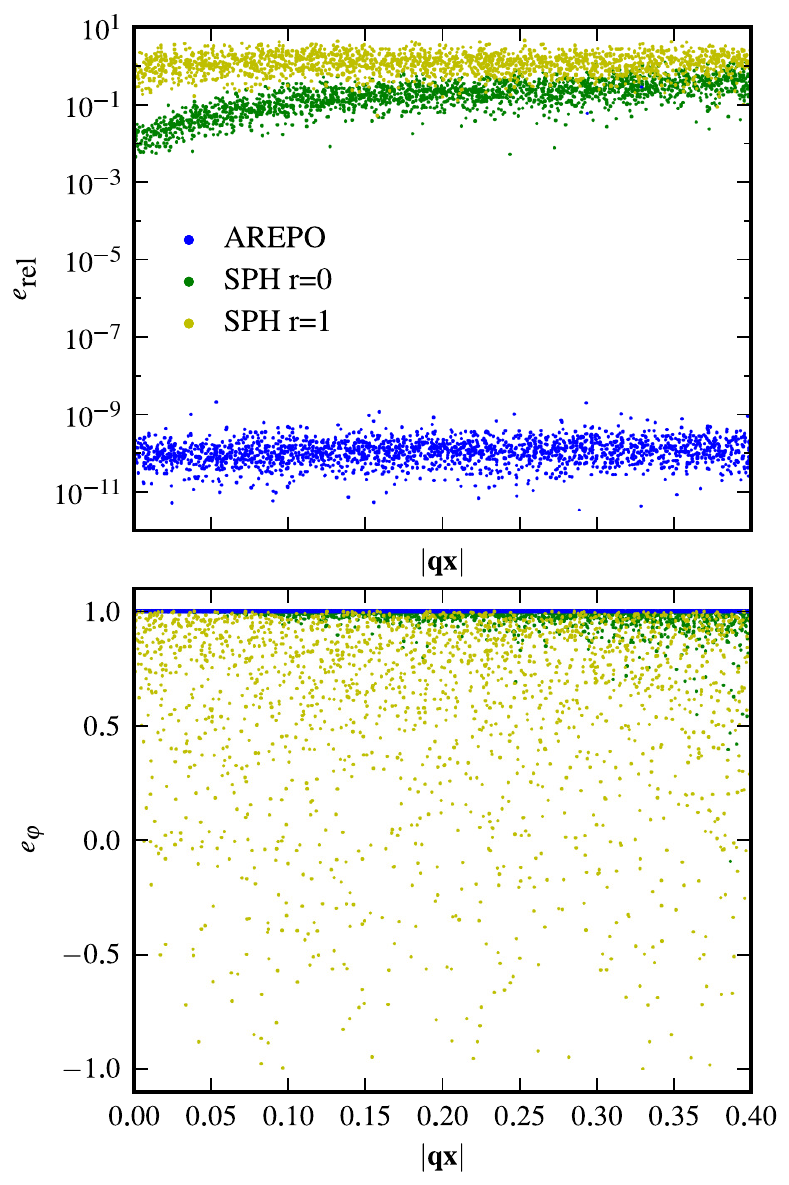}
\caption{The top panel shows a scatter plot of the relative errors of
  pressure gradient estimates in a simple test set-up. The bottom
  panel shows the corresponding errors in the direction of the
  estimated pressure gradients.  The errors of {\small AREPO} are at
  the level of machine precision.  However, SPH shows severe errors in
  the pressure gradient estimates, with a size that depends on the
  magnitude of the pressure itself. If a constant offset of $r=1$ is
  added to the pressure profile, the gradient errors are about ten
  times larger, exceeding even $100\%$. }
\label{fig:dis:err}
\end{center}
\end{figure}

In order to demonstrate this point explicitly and quantify the
typical noise in the pressure gradient estimates of SPH and {\small
  AREPO}, we have carried out a simple experiment. To this end we used
the particle coordinates $\vec{x}_i$ of the last snapshot of our S3
subsonic simulation run, which is representative for the
 particle distribution typically encountered in SPH in
this situation.  We then assigned entropies to the particles (taking
their density estimate into account) such that the pressures $P_i =
P(\vec{x}_i)$ of individual particles were given by the analytic
pressure profile
\begin{equation}
  P(\vect{x}) = P_0 \, \vect{q}\, \vect{x} + r,
\end{equation}
which is a simple linear gradient in the $\vect{q}$-direction (our
results are independent of the actual orientation of this vector) with
a constant pressure offset $r$.  The SPH estimate for the pressure
gradient was then inferred from the particle acceleration
$\vect{a}_{\textrm{SPH}}$ computed by the SPH code as
\begin{equation}
  \vect{\nabla} P = - \vect{a}_{\textrm{SPH}}\,\rho ,
\end{equation}
which is the relevant quantity that ultimately enters the discretized
equation of motion.  We can then consider the relative error of these
SPH pressure gradient estimates with respect to the known analytic
gradient.  We define the corresponding errors as
\begin{equation}
  e_{\textrm{rel}} = \frac{| \vect{\nabla}
    P-P_0\vect{q}|}{|P_0\vect{q}|} ,
  \quad \quad e_{\varphi} = \frac{  \vect{q}\cdot  \vect{\nabla} P}{|\vect{q}| | \vect{\nabla} P|} = \cos{\phi} , 
\end{equation}
and show them as scatter plots for a random subset of the points in
Figure~\ref{fig:dis:err}.

For comparison, we also carried out the equivalent procedure for the
{\small AREPO} code, based on the same particle coordinates.  The
resulting errors are also shown in Figure~\ref{fig:dis:err}. {\small
  AREPO} clearly calculates the pressure gradients highly accurately,
both in magnitude and angle. In fact, {\small AREPO}'s gradient
estimate is second-order accurate, independent of the distribution of
points \citep{Springel2010}, implying that a linear gradient should be
reproduced essentially to machine precision, which we find is also the
case here.  In contrast, SPH shows a huge scatter in both error
measures. In fact, the magnitude of the absolute error can in extreme
cases be up to twice as large as the value of the gradient itself, and
also the angular errors are significant. Furthermore, the errors rise
with the magnitude of $p$. If a pressure offset $r=1$ is added to the
pressure profile, the relative error is increased by about an order of
magnitude, which can be understood in terms of a higher $\vec{E}_0$
error in equation (\ref{eqnE0}).  We have also repeated the
calculation for increased numbers of neighbours, which reduces the
error, albeit only weakly.  We note that these large errors occur for
a rather simple problem -- a spatially constant gradient. This makes
it clear that standard SPH has comparatively low-order accuracy for
smooth flow.

We thus think that the problems of SPH in resolving subsonic
turbulence are serious. It is unlikely that they can be solved by just
increasing the resolution, reducing the artificial viscosity, or the
number of smoothing neighbours. This is because changing the
artificial viscosity parameterization does not improve the gradient
estimates, and will hence not be able to resolve the underlying
problem. Furthermore, obtaining better gradient estimates through a
larger number of SPH smoothing neighbours is efficiently blocked by
the clumping instability, at least for the ordinary kernel of classic
SPH. In any case, what appears to be needed for better results in the
subsonic regime are better gradient estimates. Some extensions and
improvements of the standard SPH formulation that go into this
direction have already been proposed \citep[e.g.][]{Price2008,
  Hess2010, Cullen2010, Read2010, Read2011, Abel2011}. It will remain
to be seen whether any of them provides a robust and generally
applicable alternative to standard SPH.
 
We should clarify that despite the large errors in gradient estimates,
it remains true that SPH has very good conservative properties. This
feature allows it to still produce physically sensible fluid behaviour
in many situations despite the subsonic noise, especially in
shock-dominated regimes where the accuracy of gradient estimates is
much less important. Our results hence justify the application of
standard SPH in studies of supersonic turbulence, provided the Mach
number is really high.  At the same time, our results raise
significant concerns for applications of SPH in regimes where subsonic
phenomena such as turbulence are important.  This is for example
expected to be the case in cosmological structure formation. Indeed,
recent studies have already presented evidence that the accuracy
problems of SPH in the treatment of the generation of turbulence and
of fluid instabilities such as the Kelvin-Helmholtz instability affect
galaxy formation directly \citep{Vogelsberger2011, Keres2011,
  Sijacki2011}.

Another important conclusion from our results is that the new
moving-mesh code {\small AREPO} is highly competitive with the
accuracy of ordinary Eulerian mesh-codes for studies of turbulence.
In the subsonic regime it produces essentially equivalent results as
ordinary Eulerian codes, with a slightly reduced dissipativeness of
the scheme. In the supersonic regime it however features a higher
effective resolution at the same number of resolution elements. In
particular, the velocity and density power spectra can be traced to
smaller scales, and there is generally less dissipation as a function
of scale due to reduced advection errors.  If just pure hydrodynamics
without self gravity is considered, a moving-mesh calculation with
{\small AREPO} is however more costly than a corresponding fixed-mesh
or SPH calculation with the same number of resolution elements. It is
clear that an accurate description of turbulent gas motions is highly
desirable for a versatile astrophysical code, and we have shown here
that {\small AREPO} is able to meet these requirements.

\section*{Acknowledgements}

We would like to thank Christoph Federrath and Daniel Price for their
turbulent driving routine, which we have incorporated in modified form
in our {\small AREPO} and {\small GADGET-3} codes. We would also like
to thank Christoph Federrath, Dusan Keres, Mordecai-Mark Mac Low,
Colin McNally for very insightful comments. AB gratefully acknowledges
financial support from the Klaus Tschira Foundation, and VS from DFG
Research Center SFB 881, `The Milky Way System'.

\bibliographystyle{mn2e} 
\bibliography{paper}

\end{document}